\documentclass[aps, 10pt,pra,twocolumn,groupedaddress]{revtex4-1}

\usepackage{graphicx}
\usepackage{dcolumn}
\usepackage{bm}
\usepackage{epsfig}
\usepackage{amsmath}
\usepackage{amssymb}
\usepackage{color}
\usepackage{bbm}
\usepackage{braket}
\usepackage{float}
\usepackage{enumitem}
\usepackage{tensor}
\usepackage{graphicx}
\usepackage{xcolor}
\usepackage[english]{babel}
\usepackage{bm,graphicx,url,epsf,color}
\usepackage{amssymb}
\usepackage{verbatim}
\usepackage[commandnameprefix=always]{changes}
\usepackage{amsthm}
\usepackage[normalem]{ulem}
\usepackage{changes}
\usepackage[dvipsnames]{xcolor}
\usepackage{overpic}
\usepackage{physics}

\usepackage{tikz}
\newcommand*\circled[1]{\tikz[baseline=(char.base)]{
            \node[shape=circle,draw,inner sep=1pt] (char) {#1};}}

\newcommand{\stkout}[1]{\ifmmode\text{\sout{\ensuremath{#1}}}\else\sout{#1}\fi}

\usepackage[colorlinks=true,citecolor=blue,linkcolor=blue,urlcolor=blue]{hyperref}

\begin{document}

\title{Electrostatics-induced breakdown of the integer quantum Hall effect in cavity QED}
\author{Gian Marcello Andolina$^{1}$}
\altaffiliation{These authors contributed equally to this work.}
\author{Zeno Bacciconi$^{2}$}
\altaffiliation{These authors contributed equally to this work.}
\author{Alberto Nardin$^{3}$}
\altaffiliation{These authors contributed equally to this work.}
\author{Marco Schirò$^{1}$}
\author{Peter Rabl$^{4,5,6}$}
\author{Daniele De Bernardis$^{7,8}$}
\email{daniele.debernardis@cnr.it}
\altaffiliation{Corresponding author}

\affiliation{$^1$ JEIP, UAR 3573 CNRS, College de France, PSL Research University,
11 Place Marcelin Berthelot, F-75321 Paris, France}
\affiliation{$^2$SISSA-International School for Advances Studies, via Bonomea 265, 34136 Trieste, Italy
ICTP-The Abdus Salam International Centre for Theoretical Physics,
Strada Costiera 11, 34151 Trieste, Italy}
\affiliation{$^3$Universite Paris-Saclay, CNRS, LPTMS, 91405, Orsay, France}
\affiliation{$^4$Walther-Mei\ss ner-Institut, Bayerische Akademie der Wissenschaften, 85748 Garching, Germany}
\affiliation{$^5$Technische Universit\"at M\"unchen, TUM School of Natural Sciences, Physics Department, 85748 Garching, Germany} 
\affiliation{$^6$Munich Center for Quantum Science and Technology (MCQST), 80799 Munich, Germany} 
\affiliation{$^7$Istituto Nazionale di Ottica (CNR-INO), c/o LENS via Nello Carrara 1, Sesto F.no 50019, Italy}
\affiliation{$^8$European Laboratory for Non-Linear Spectroscopy (LENS), Via Nello Carrara 1, Sesto Fiorentino 50019, Italy}

\date{\today}

\begin{abstract}
We address the prevailing theoretical explanation of the recently observed breakdown of the integer quantum Hall effect in a two-dimensional electron gas embedded in a metallic split-ring resonator. Within the same single-particle description of quantized Hall conductance, we compare previously proposed vacuum-induced transport modifications against an alternative mechanism that explains this breakdown in terms of non-chiral edge channels arising solely from electrostatic boundary effects. This direct comparison shows that for experimentally relevant parameters, the electrostatic contribution exceeds that of any vacuum-induced conductance modifications by many orders of magnitude and yields characteristic transport signatures and energy scales that align well with experimental observations. This finding sheds new light on this puzzling phenomenon, supporting an electrostatic rather than a vacuum-related interpretation that can be directly tested in experiments.
\end{abstract}

\maketitle

The integer quantum Hall (IQH) effect, discovered by von Klitzing in 1980 \cite{klitzing_new_1980}, represents a prototypical example of topological physics that emerges in a two-dimensional electron gas (2DEG) in the presence of a strong perpendicular magnetic field. It describes the quantization of the Hall conductivity in integer multiples of $e^2/h$, which remains remarkably robust against material imperfections and electron-electron interactions, due to the topological nature of electronic wavefunctions.
However, recent experiments reported a significant distortion of the Hall plateaus once the 2DEG is placed inside a THz split-ring resonator \cite{appugliese_breakdown_2022, enkner_tunable_2025}. 
To explain these puzzling observations, various mechanisms that emerge from the strong coupling of electrons to the vacuum fluctuations of the quantized resonator field have been proposed. These include long-range electron hoppings mediated by vacuum cavity fluctuations~\cite{ciuti_cavity-mediated_2021,arwas_quantum_2023, borici_cavity-modified_2024}, a renormalization of the $g^\star$-factor by cavity vacuum effects \cite{enkner_tunable_2025} and lower-energy polaritonic states that worsen the temperature resilience~\cite{rokaj_weakened_2023} and contribute to finite frequency resistivity \cite{cardoso2025cavityquantumhallhydrodynamics, yang2025quantumhalleffectchiral}.
In the absence of compelling alternative explanations, the physical picture of a vacuum-induced breakdown of the Hall effect has consequently become the prevailing interpretation of this and related cavity-induced transport phenomena.

In this letter, we revisit the physics of Hall transport inside a split-ring resonator by taking both vacuum-induced and electrostatic boundary modifications into account. 
As already emphasized in previous theoretical works~\cite{de_bernardis_cavity_2018,saez-blazquez_can_2023,Amelio2021,Kotov2025,SanchezMartinez2024,Pantazopoulos2024,FernandezDeLaPradilla2025,andolina_quantum_2025,andolina_amperean_2024}, electrostatic effects are expected to exceed quantum vacuum contributions in magnitude under most conditions. This raises important questions about the role of electrostatic effects in cavity-modified transport phenomena~\cite{paravicini-bagliani_magneto-transport_2019,appugliese_breakdown_2022, enkner_testing_2024, enkner_tunable_2025, graziotto_cavity_2025}, which have received little attention so far, but may drastically change our microscopic understanding of these experimental results. Based on a single-particle picture of the integer quantum Hall effect \cite{Laughlin_PhysRevB.23.5632, Thouless_PhysRevB.27.6083, Huckestein_RevModPhys.67.357, Thouless_PhysRevLett.49.405, halperin_quantized_1982, Hatsugai_PhysRevB.48.11851, trugman_localization_1983, macdonald1994introductionphysicsquantumhall,tong_lectures_2016}, in this paper we identify a specific mechanism that leads to the breakdown of conductance quantization in a cavity quantum Hall setup and arises from purely static image charges induced on the electrodes of the metallic split-ring resonator. Moreover,  within the same formalism and under the same experimental conditions, we find that the corresponding vacuum-induced modifications of the quantum Hall system are several orders of magnitude weaker.

While based on a simplified single-electron description, this analysis illustrates the vastly different roles of electrostatic and vacuum-induced effects in a cavity quantum Hall setting. 
For systems where screening is negligible, it also provides quantitative predictions for transport modifications that can be tested against alternative theoretical models. 
In heterostructures, as used in Refs. \cite{paravicini-bagliani_magneto-transport_2019,appugliese_breakdown_2022, enkner_testing_2024, enkner_tunable_2025, graziotto_cavity_2025}, accurate quantitative predictions still require a fully self-consistent treatment of electron-electron interactions and a detailed modeling of the electrode geometry, which may alter the precise mechanism behind electrostatic backscattering channels. 
However, even substantial interaction effects are unlikely to significantly alter the relative strengths of vacuum-induced and electrostatic boundary effects, suggesting that the latter remains the dominant source of perturbation in a typical cavity quantum Hall setup \cite{debernardis_in_preparation}. 
Beyond transport, this observation is highly relevant for developing refined microscopic understanding of a much broader class of cavity-modified quantum materials \cite{schlawin_cavity_2022,garcia-vidal_manipulating_2021,ashida_quantum_2020,Schuler_Vacuum_2020,andolina_amperean_2024,passetti_cavity_2023,bacciconi_first-order_2023,dmytruk_gauge_2021,latini_cavity_2019,bacciconi_majoranaprb_2024, de_bernardis_relaxation_2023,chiriaco_critical_2022, de_bernardis_light-matter_2023,dmytruk_controlling_2022,Bacciconi_rydbergcavity_2025prl,bacciconi_dissipation_2025,schachenmayer2015cavity,hagenmuller2017cavity}, which will eventually allow us to isolate genuine vacuum corrections in these systems from other, more mundane effects.

\begin{figure}
    \centering
    \includegraphics[width=\columnwidth]{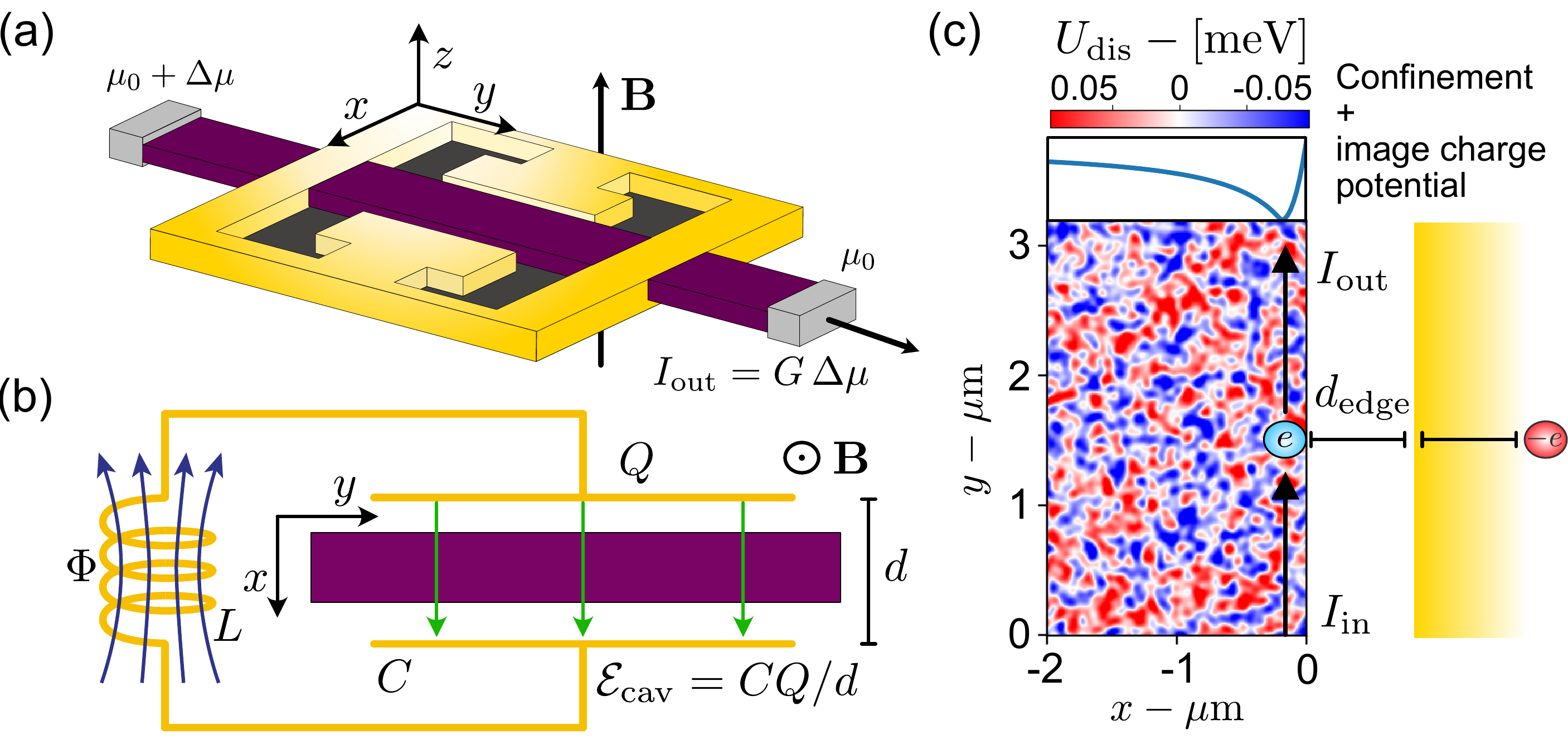}
    \caption{(a) Sketch of a cavity quantum Hall system, where a single electron in an external magnetic field $\textbf{B}$ is strongly coupled to the quantized field $\mathcal{E}_{\rm cav}$ of a split-ring resonator. 
    (b) The resonator is modeled as a lumped-element $LC$ circuit with the Hall bar inside the two capacitor plates separated by a distance $d$.  (c) Example of the disorder potential $U_{\rm dis}({\bf r})$ with a sketch of the combined potentials $U_{\rm conf}({\bf r})+U_{\rm im}({\bf r})$ (upper panel, blue solid line). 
 }
    \label{fig:1}
\end{figure}

{\it Quantum Hall system in a cavity.}---For the following analysis, we consider a Hall bar of dimensions $L_x$ and $L_y$ embedded in a metallic split-ring resonator (see Fig.~\ref{fig:1}).
Following previous literature on cavity quantum Hall systems \cite{ciuti_cavity-mediated_2021, arwas_quantum_2023, borici_cavity-modified_2024}, we neglect screening and electron-electron interaction effects and focus purely on the physics of a single electron subjected to a strong magnetic field $B$ perpendicular to the $x-y$ plane and moving close to the metallic plates of a cavity, as exemplified in Fig.~\ref{fig:1}(a,c). 
Within this single-electron picture \cite{Laughlin_PhysRevB.23.5632, Thouless_PhysRevB.27.6083,tong_lectures_2016}, the Fermi energy $E_F$ parametrizes the energy at which the single electron is injected in the Hall bar; in the standard scenario, i.e. in the absence of the metallic resonator electrode, the 2D electron system displays a quantized electric conductance $G=(e^2/h) (1+\ell)$, where $\ell =0,1,2,3,\dots$ is the Landau level (LL) principal quantum number with energy $E_{\ell}$ below the Fermi energy, $E_{\ell}<E_F$ \cite{arwas_quantum_2023}. 
This effect can be attributed to the presence of chiral edge modes \cite{buttiker_absence_1988,tong_lectures_2016}, and it is thus 
protected \cite{von_klitzing_40_2020} against local disorder. 

The single particle Hamiltonian describing the electron's motion is given by
\begin{equation}\label{eq:ham_1}
    H_e = \frac{\left[\mathbf{p}+e\mathbf{A}_{\rm ext}(\mathbf{r})\right]^2}{2m} + U(\textbf{r}).
\end{equation}
Here, $\mathbf{p} = (p_x, p_y)$ and $\mathbf{r} = (x, y)$ denote the momentum and position operators of the electron with mass $m$ and charge $e$. The external vector potential is taken in Landau gauge, $\mathbf{A}_{\rm ext}(\mathbf{r}) = (0,xB)$.
In our minimal setup, the single-particle potential 
\begin{equation}\label{eq:pot_Efield}
U(\mathbf{r}) = U_{\rm dis}(\mathbf{r}) + U_{\rm im}(\mathbf{r}) + U_{\rm conf}(\mathbf{r})
\end{equation}
includes three contributions: a disorder potential, an image-charge contribution due to the nearby metallic boundaries, and a confining potential. 

The disorder potential $U_{\rm dis}(\mathbf{r})$ arising from material imperfections is fundamental for the modeling of transport properties of IQH systems~\cite{ando_electronic_1982, ando_localization_1985, tong_lectures_2016}. 
For concreteness, we assume a Gaussian-correlated random potential~\cite{ando_electronic_1982,de_bernardis_magnetic-field-induced_2022,borici_cavity-modified_2024}, which is characterized by
\begin{equation}\label{eq:disorder_pot}
\overline{ U_{\rm dis}(\mathbf{r}) U_{\rm dis}(\mathbf{r}') } = E_{\rm dis}^2  e^{-|\mathbf{r}-\mathbf{r}'|^2/\xi_c^2}.
\end{equation}
Here, $\overline{\cdots}$ denotes the average over disorder realizations, $E_{\rm dis}\ll \hbar\omega_B$ sets the disorder strength, and $\xi_{c}$ is the spatial correlation length. A representative realization of $U_{\rm dis}(\mathbf{r})$, as used in our numerical simulations below, 
is shown in Fig.~\ref{fig:1}(c). 

The image charge potential $U_{\rm im}(\mathbf{r})$ arises from the image of the transported electron in the metallic cavity electrode near the conducting edge, as depicted in Fig. \ref{fig:1}(c).
While a fully accurate description of $U_{\rm im}(\mathbf{r})$ would require a detailed electrostatic modeling of the split-ring geometry, a proper inclusion of interparticle interactions and other experimental details ~\cite{chklovskii_electrostatics_1992,chklovskii_ballistic_1993,siddiki_incompressible_2004,lier_self-consistent_1994,Oh_PhysRevB.56.13519}, here we simply retain the effect of a single metallic boundary near the conducting edge. This simplification permits a direct comparison with equivalent single-particle models of vacuum-fluctuations~\cite{ciuti_cavity-mediated_2021,arwas_quantum_2023, borici_cavity-modified_2024} and establishes a clear hierarchy in the order of magnitude  between the two effects.
Specifically, we model the region $x \geq d_{\rm edge}$ as a perfect conductor, while electrons are free to move in the half-space $x \leq 0$. An electron contributing to transport thus interacts with its electrostatic image that is located at a distance $2|x - d_{\rm edge}|$ across the metallic plane. This yields an attractive Coulomb potential of the form
\begin{equation}\label{eq:single_plate_image_pot}
    U_{\rm im}({\bf r}) = -\frac{e^2}{8\pi\epsilon| x-d_{\rm edge}|}, 
\end{equation}
where $\epsilon\approx 13\epsilon_0$ is the dielectric constant of GaAs. 
We remark that assuming an infinite conducting plane is a crude approximation, but it captures the essential feature of in-plane variation of the electric field. 

The confinement 
$U_{\rm conf}(\mathbf{r})$ is a smooth potential over the magnetic length scale $l_B \equiv (\hbar/eB)^{1/2}$. 
If sufficiently steep, its combination with the image charge potential $U_{\rm im}({\bf r}) + U_{\rm conf}(\mathbf{r})$ gives rise to a non-monotonous shape, attractive toward the edge and with a local minimum, as shown in Fig.~\ref{fig:1}(c) and Fig.~\ref{fig:2}(a).
As we will see in the following, this \emph{pocket potential} can lead to the breakdown of conductance quantization~\cite{Silvestrov_PhysRevB.77.155436, akiho_counterflowing_2019, zeldov_2019, zeldov_2021, Moreau_2021}.

\begin{figure}
    \centering
    \includegraphics[width=\columnwidth]{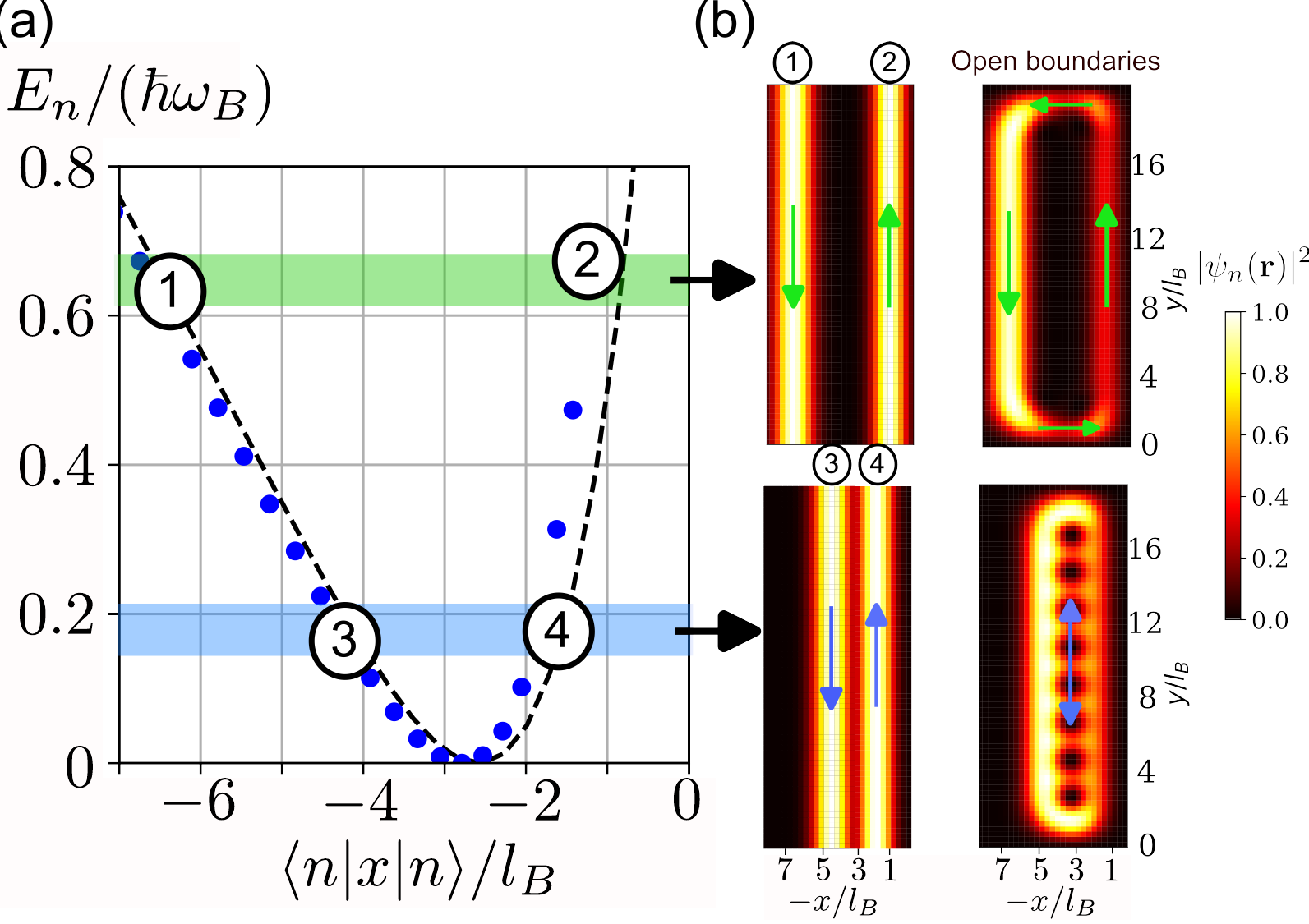}
    \caption{
    (a) Spectrum of the LLL against eigenstate position $\langle{n|x|n\rangle}$ close to the edge and distorted by an external potential $U(x) = Ax^6 - e\mathcal{E}_{\rm ext} x$ mimicking the action of the confinement and image charge potentials. 
    Here $e\mathcal{E}_{\rm ext} = \Gamma / l_B$.
    (b) Wavefunction $|\psi_n(\mathbf{r})|^2$ as a function of position for the marked eigenstates. With PBC on the left, with open boundary condition (OBC) on the right. Parameters in the End Matter (EM).
    }
    \label{fig:2}
\end{figure}

{\it Breakdown of conductance quantization.}---We first analyze a minimal mechanism that captures the loss of Hall conductance quantization. For simplicity, we assume that the confining potential is translationally invariant along the $y$-direction, such that $U_{\rm conf}(\mathbf{r}) = U_{\rm conf}(x)$, and we neglect disorder by setting $E_{\rm dis} = 0$. Under these assumptions, the total potential $U(x)$ develops a single, well-defined minimum near the system boundary at $x \approx 0$. This effective pocket potential is illustrated schematically in Fig.~\ref{fig:1}(c). Assuming that $U(x)$ is sufficiently smooth on the scale of the magnetic length $l_B$, the energies of the Landau levels follow adiabatically the shape of the potential, $E_\ell(x) \approx \hbar \omega_B\, \ell + U(x)$, where $\omega_B = eB/m$ is the cyclotron frequency. 

Consider now a pair of eigenstates at the same energy $E_1=E_2=E$ whose wavefunctions are centered around two points $x_1 > x_2$ on opposite sides of the pocket [for example, the pair of modes  \protect\circled{1}, \protect\circled{2} and \protect\circled{3}, \protect\circled{4} in Fig. \ref{fig:2}]. Along the $y$-direction, these wavefunctions are delocalized plane waves with a velocity $v_y\propto\partial_x U(x)$ \cite{tong_lectures_2016}. Importantly, states located at opposite sides of the pocket propagate in opposite directions along the equipotential lines of $U(x)$. As long as the separation between such states remains large, they represent two independent chiral conduction channels. However, when $|x_1-x_2|\lesssim l_B\sqrt{\ell+1}$, any perturbation that breaks momentum conservation along the $y$ direction couples the two modes and can lead to backscattering of injected electrons \cite{Moreau_2021, moreau_contacts_2021}. Intuitively, this can be anticipated by the formation of non-chiral modes with reduced conductivity. This is exemplified in Fig.~\ref{fig:2}(b), where we show that in the case of open boundary conditions, the nearby channels \protect\circled{3} and \protect\circled{4} hybridize and form a standing wave \cite{de_bernardis_cavity_2018}. This is not the case for the well-separated modes \protect\circled{1} and \protect\circled{2}.

For the image-charge-induced pocket potential of interest, the formation of those non-chiral and non-conductive standing-wave modes affects only the lowest energy states of each LL, up to an energy that is roughly determined by the condition $|x_1-x_2|= l_B\sqrt{\ell+1}$. This allows us to introduce the characteristic \emph{backscattering threshold energy} 

\begin{equation}\label{eq:tunneling_energy}
    \Gamma_\ell = l_B\sqrt{\ell+1}\,\partial_{x} U_{\rm im}\vert_{x=0} =\frac{e^2}{8\pi \epsilon } \frac{l_B\sqrt{\ell+1}}{d_{\rm edge}^2}.
\end{equation}
When the Fermi energy lies just at the bottom of the pocket, in an energy window of order $\sim\Gamma_{\ell}$, we expect a nonquantized Hall conductance, since an injected electron can backscatter.

\begin{figure}
    \centering
    \includegraphics[width=\columnwidth]{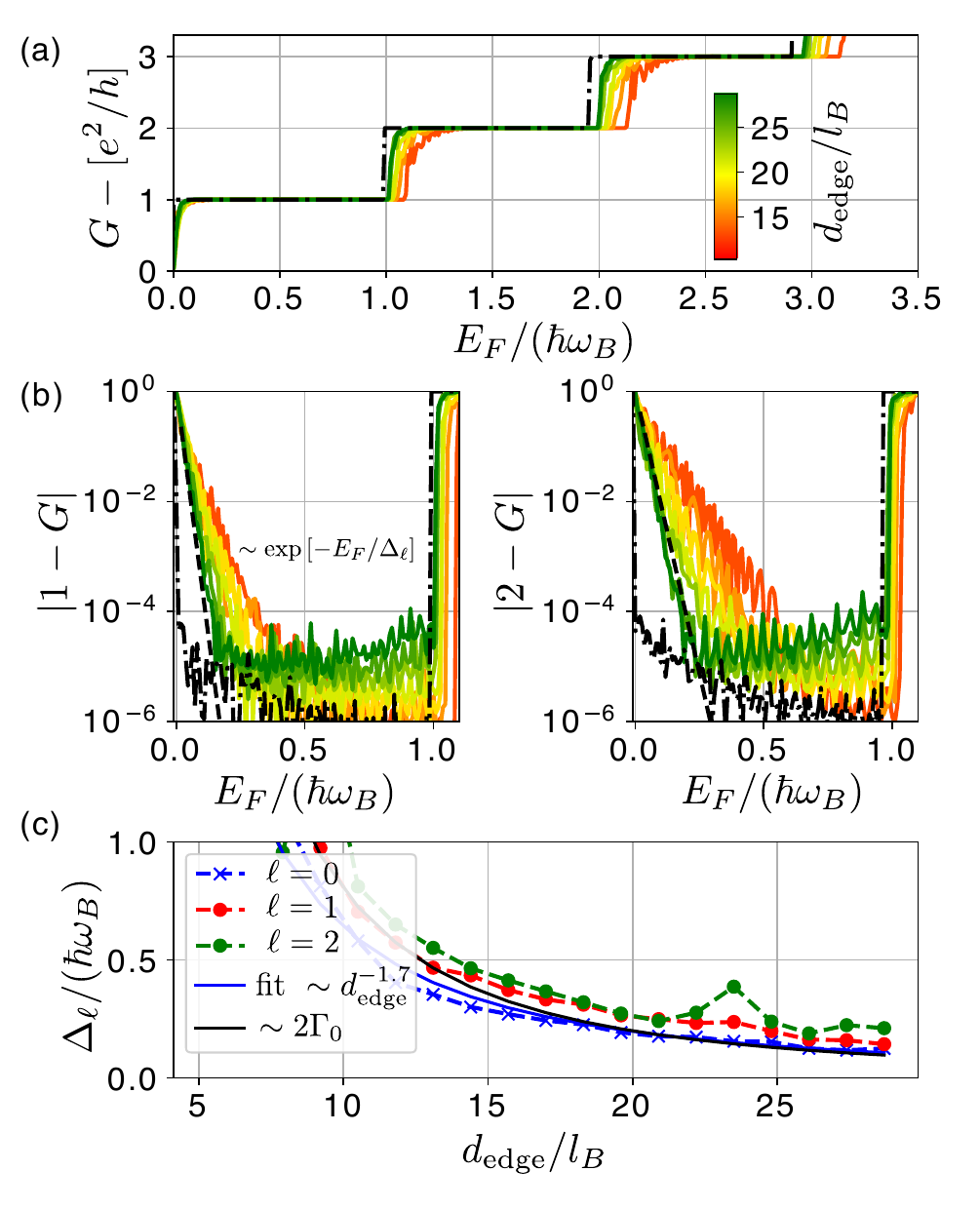}
    \caption{(a) Conductance $G(E_F)$ as a function of the Fermi energy, for a reference system (black dashed) and a system with a metallic plate at distance $d_{\rm{edge}}/l_B=3.5-30$ (red-green), that is including the image charge potential in Eq. \eqref{eq:single_plate_image_pot}. 
    (b) Deviation from quantization of the conductance at $\ell=0$ (left) and $\ell=1$ (right) plateaus. The dashed line is an exponential fit of the curve at $d_{\rm edge}/l_B = 30$, using the fitting function  $f_{\rm fit}(E_F|\Delta , A)=\exp[ - E_F/(\hbar \Delta_{\ell}) + A ]$ with $\Delta_{\ell}$, $A$ as free parameters. 
    (c) Extracted exponential slope $\Delta_{\ell}$ as a function of the plate distance $d_{\rm edge}$. The black solid line is the scaling of $2\Gamma$, the blue solid line is a numerical fit over the curve at $\ell=0$, indicating a scaling $\sim d_{\rm edge}^{-1.7}$. 
    All the conductance curves are shifted such that the minimum energy eigenvalue is at 0.
    Parameters can be found in the EM, they realize $\Gamma_{\ell=0}/\hbar \omega_B \sim 0.4$ at $d_\mathrm{edge}/l_B=10$.}
    \label{fig:3}
\end{figure}

\emph{Numerical simulations.}---In order to support the picture provided above in the single-particle model, we perform transport simulations using the library \texttt{Kwant} \cite{groth_kwant_2014, arwas_quantum_2023} in the continuum limit of the Harper-Hofstadter model, which is well-suited to describe LL physics (see~\cite{SM} for details). We use the scattering method, where the transmission coefficient is evaluated for a ballistic electron injected just above the Fermi energy $E_F$. This transmission coefficient corresponds to a two-terminal conductance $G(E_F)$, which is quantized in an IQH system whenever $E_F$ lies between two LL, effectively counting the number of chiral edge states \cite{beenakker_quantum_1991}.

In Fig.~\ref{fig:3} we show numerical results for the conductance $G(E_F)$ at zero temperature, $T=0$, comparing the case without (black dashed lines) and with (colored lines) the image charge potential, for a range of plate distance $d_{\rm{edge}}/l_B$. Large plateaus are visible in both cases [see Fig.~\ref{fig:3}(a)] and extend over almost all values of $E_F$,  except for small windows of width $\Delta_\ell$ above $E_\ell$, 
where a breakdown of the quantized conductance is observed for $U_{\rm{im}}\neq 0$. For the lowest LL, we fit the deviations from exact quantization with an exponential curve [see Fig.~\ref{fig:3}(b)], which can be motivated within an approximate semiclassical treatment of the problem~\cite{SM} and allows us to extract the energy width $\Delta_\ell$. We plot this as a function of the plate distance $d_{\rm{edge}}$ in Fig.~\ref{fig:3}(c). 
For $\ell=0$ this curve is fitted by a power-law in $d_{\rm{edge}}$ and compared to the scaling $\Delta_{\ell=0} = 2\Gamma_0\propto d_{\rm{edge}}^{-2}$. 
This comparison confirms the intuitive picture developed above. It shows that the backscattering threshold energy $\Gamma_\ell$ and its dependence on $d_{\rm{edge}}$ capture the relevant energy scale that determines the breakdown of quantized transport very accurately. 
A more in-depth and accurate analysis of the nonquantized transport is given in~\cite{SM} and more detailed simulations that also include the spin degree of freedom are shown in End Matter.

\emph{Parameter estimates}.---We now discuss the relevance of the electrostatic mechanism explained so far for cavity QED quantum Hall experiments. 
The important parameter emerging from our toy model is the strength of the image-charge potential in-plane gradient, parametrized by $\Gamma_\ell$ [see Eq.~\eqref{eq:tunneling_energy}], and the energy separation of the LLs. The latter is given by the cyclotron energy $\hbar\omega_B$ or, including the spin degree of freedom, the Zeeman energy gap $E_Z$ (see End Matter for further details). 
While a precise estimate for $\Gamma_\ell$ would require a detailed analysis of the split-ring resonator's sharp structure, a meaningful lower-bound estimate can be provided by using the infinite plate approximation and the distance $d_{\rm{edge}}\sim 200\,$nm estimated for the main sample of Ref.~\cite{appugliese_breakdown_2022}. 
Together with the other parameters $\ell=4$, $B=1\,$T and $E_Z=0.2\,$meV, we obtain an estimate $\Gamma_{\ell=4}/(\hbar \omega_B)\approx 4\%$ valid for even plateaus and $\Gamma_{\ell=4}/E_Z\approx 38\%$ for odd plateaus. 
With a minimal further adjustment of the parameters, a full breakdown of conductance quantization in the odd plateau, $\Gamma_{\ell=4}/E_Z\approx 100\%$, can be reached.

\emph{Vacuum field contribution}.---In addition to electrostatic effects, the transport of electrons is influenced by their coupling to the quantized field of the split-ring resonator~\cite{ciuti_cavity-mediated_2021, arwas_quantum_2023, borici_cavity-modified_2024}. 
We model the cavity as a lumped-element resonator~\cite{de_bernardis_hybrid_2025,cardoso2025cavityquantumhallhydrodynamics} with inductance $L$, capacitance $C$, resonance frequency $\omega_{LC} = 1/\sqrt{LC}$, and an approximately homogeneous electric field between the capacitor plates.  
The total single-electron Hamiltonian is then given by~\cite{de_bernardis_cavity_2018, saez-blazquez_can_2023, andolina_quantum_2025}
\begin{equation}\label{eq:ham_tot}
     H= H_e +  \frac{\left(Q - Q_e \right)^2}{2C} + \frac{\Phi^2}{2L},
\end{equation}
where $\Phi$ and $Q$ are the quantized flux and charge operators, which obey $[\Phi,Q]=i\hbar$. The offset $Q_e = e(x/d)$ represents the amount of charge that is induced by the electron at position $x$ across the capacitor plates that are separated by distance $d$~\cite{de_bernardis_cavity_2018, saez-blazquez_can_2023}. This leads to a dipole-type coupling between the electron displacement $x$ and the quantized electric cavity field $E_c =Q/(cd)=\mathcal{E}_{\rm vac}(a+a^\dagger)$.

When the resonator frequency is off-resonant with respect to the characteristic LL energy scales, $\omega_{LC} \neq \omega_B$, we can use second-order perturbation theory to eliminate the cavity mode and derive an effective Hamiltonian for the electron only. This corresponds to a generalized Lamb shift for extended LL, which was first investigated in Ref.~\cite{ciuti_cavity-mediated_2021}. As detailed in~\cite{SM}, the relevant vacuum-induced correction is given by, 
\begin{equation}\label{eq:Hpa_main}
   H_{\rm vac}^{(2)} = -\frac{(e\mathcal{E}_{\rm vac}l_B)^2}{2\hbar\omega_{LC}}\frac{\omega_B}{\omega_{LC}} \mathcal{K},
\end{equation}
where $\mathcal{K}$ is the dimensionless, single-electron,  kinetic energy operator. While in disordered systems, $\mathcal{K}$ can induce transitions between localized LL in the bulk or between edge modes~\cite{ciuti_cavity-mediated_2021,arwas_quantum_2023}, its matrix elements are $\sim O(1)$, independently of the degeneracy of the LLs~\cite{SM}. 
Thus, the potential impact of this vacuum-induced correction can be estimated from the energy scale determined by the prefactor in Eq.~\eqref{eq:Hpa_main}. 

Using the parameters given in Tabs.~\ref{tab:electrons}-\ref{tab:resonator}, we obtain vacuum-field strengths in the range of $\mathcal{E}_{\rm vac} \sim 1-10$V/m and
\begin{equation}
    \frac{(e\mathcal{E}_{\rm vac}l_B)^2}{2\hbar\omega_{LC}}\frac{\omega_B}{\omega_{LC}} \sim 10^{-9}-10^{-7}{\rm meV}.
\end{equation}
Therefore, for realistic device parameters, the expected, purely vacuum-induced corrections are in the range of $10^{-9}$–$10^{-7}$ meV, which is several orders of magnitude below the electrostatic backscattering scale identified above. It is worth noticing that in a many-electron system, the resonant optical response \cite{ciuti_quantum_2005} is collectively enhanced by the Landau-level degeneracy $N_{\rm L}\sim10^6$~\cite{hagenmuller_ultrastrong_2010,SM}. However, such an enhancement does not appear in the single-particle vacuum-induced corrections relevant for DC Hall transport~\cite{SM}.

\emph{Discussion and conclusions.}---In summary, we have proposed and analyzed a purely electrostatic mechanism that predicts the breakdown of the quantized Hall conductance in cavity-embedded systems. To do so, we have explored how the presence of lateral unbiased metallic plates on the side of the Hall bar affects ballistic transport in the integer quantum Hall regime. Within our effective single-particle description, a non-chiral channel forms close to the edge, thereby contributing to a non-quantized response. By estimating the relevant energy scales, we find that this mechanism becomes important in experimental cavity QED systems when the spacing between the sample edge and the electrodes is on the order of or below $d_\mathrm{edge}\sim 200\,$nm.
The proposed mechanism is local in space and does not rely on any coupling between the two counter-propagating edges of the sample. These predictions can thus be readily tested against alternative vacuum-induced processes, for which we predict a negligible effect on transport under the same experimentally relevant conditions.   

While the present analysis permits a direct comparison between purely electrostatic and quantum dynamical effects in a quantum Hall transport scenario, we emphasize that the formation of a pocket potential near the sample edge arises from a single-particle picture. 
Similar electrostatic mechanisms have been discussed and observed in graphene quantum Hall setups \cite{zeldov_2019,zeldov_2021,Moreau_2021,moreau_contacts_2021}, where image charges and edge reconstruction play a key role \cite{Silvestrov_PhysRevB.77.155436} and where the current theoretical framework would be directly applicable. 
In other systems, which include the experimental setups in Refs. \cite{paravicini-bagliani_magneto-transport_2019, appugliese_breakdown_2022, enkner_testing_2024, enkner_tunable_2025, graziotto_cavity_2025}, such a single-particle picture is clearly incomplete because it neglects the self-consistent electrostatics generated by electron-electron and electron-donor interactions \cite{chklovskii_electrostatics_1992, chklovskii_ballistic_1993}. 
These interactions strongly reshape the local density and potential landscape, leading to compressible and incompressible regions that dominate the actual current distribution and transport properties of realistic Hall bars \cite{siddiki_incompressible_2004}.
A full quantitative comparison with those experiments \cite{paravicini-bagliani_magneto-transport_2019, appugliese_breakdown_2022, enkner_testing_2024, enkner_tunable_2025, graziotto_cavity_2025,xue2025observationcavitymediatednonlinearlandau} requires many-body methods such as the self-consistent Hartree framework \cite{lier_self-consistent_1994, Oh_PhysRevB.56.13519, siddiki_incompressible_2004, gerhardts_self-consistent_2020}, or more sophisticated Kohn-Sham/DFT calculations \cite{halperin_PhysRevB.49.1862, stoof_density-functional_1995,ihnatsenka_spatial_2006, ihnatsenka_spin_2006}, together with a realistic modeling of the heterostructure and the electrode's multi-terminal geometry \cite{ciuti2026commentelectrostaticsinducedbreakdowninteger,borici_cavity-modified_2024}. A considerably refined analysis that addresses these issues will be presented elsewhere \cite{debernardis_in_preparation}, but doesn't change the overall conclusion on the electrostatic nature of the cavity-induced breakdown of the quantum Hall effect.

\emph{Acknowledgements.}---We are grateful to Cristiano Ciuti, Jerome Faist and Giacomo Scalari for numerous illuminating discussions and comments on the manuscript.
We also thank Iacopo Carusotto, Raffaele Colombelli, Lorenzo Graziotto, Bianca Turini, Jules Sueiro, Dalin Borici, Enrico Di Benedetto, Simone De Liberato, Luca Giacomelli, Mohammad Hafezi, Paolo De Natale, Tecla Gabrielli, Francesco Cappelli, Cristina Rimoldi, Stefan Ludwig for critical reading of the manuscript, insights,  and very stimulating discussions. D.D.B. acknowledges funding from the European Union - NextGeneration EU, "Integrated infrastructure initiative in Photonic and Quantum Sciences" - I-PHOQS [IR0000016, ID D2B8D520, CUP B53C22001750006]. This research is part of the Munich Quantum Valley, which is supported by the Bavarian state government with funds from the Hightech Agenda Bayern Plus. G.M.A acknowledge funding from the European Union’s Horizon 2020 research and innovation programme under the Marie Sklodowska-Curie (Grant Agreement No. 101146870 – COMPASS). G.M.A and M.S. acknowledge funding from the European Research Council (ERC) under the European Union’s Horizon 2020 research and innovation programme (Grant agreement No. 101002955 – CONQUER).
 A.N. is supported by the ANR project LOQUST ANR-23-CE47-0006-02.

\bibliographystyle{mybibstyle}
\bibliography{references}

\clearpage
\newpage

\onecolumngrid

\vspace*{1cm}
\begin{center}
\textbf{\large End Matter}
\end{center}
\vspace{0.5cm}
\twocolumngrid

\begin{figure}
    \centering
    \includegraphics[width=\columnwidth]{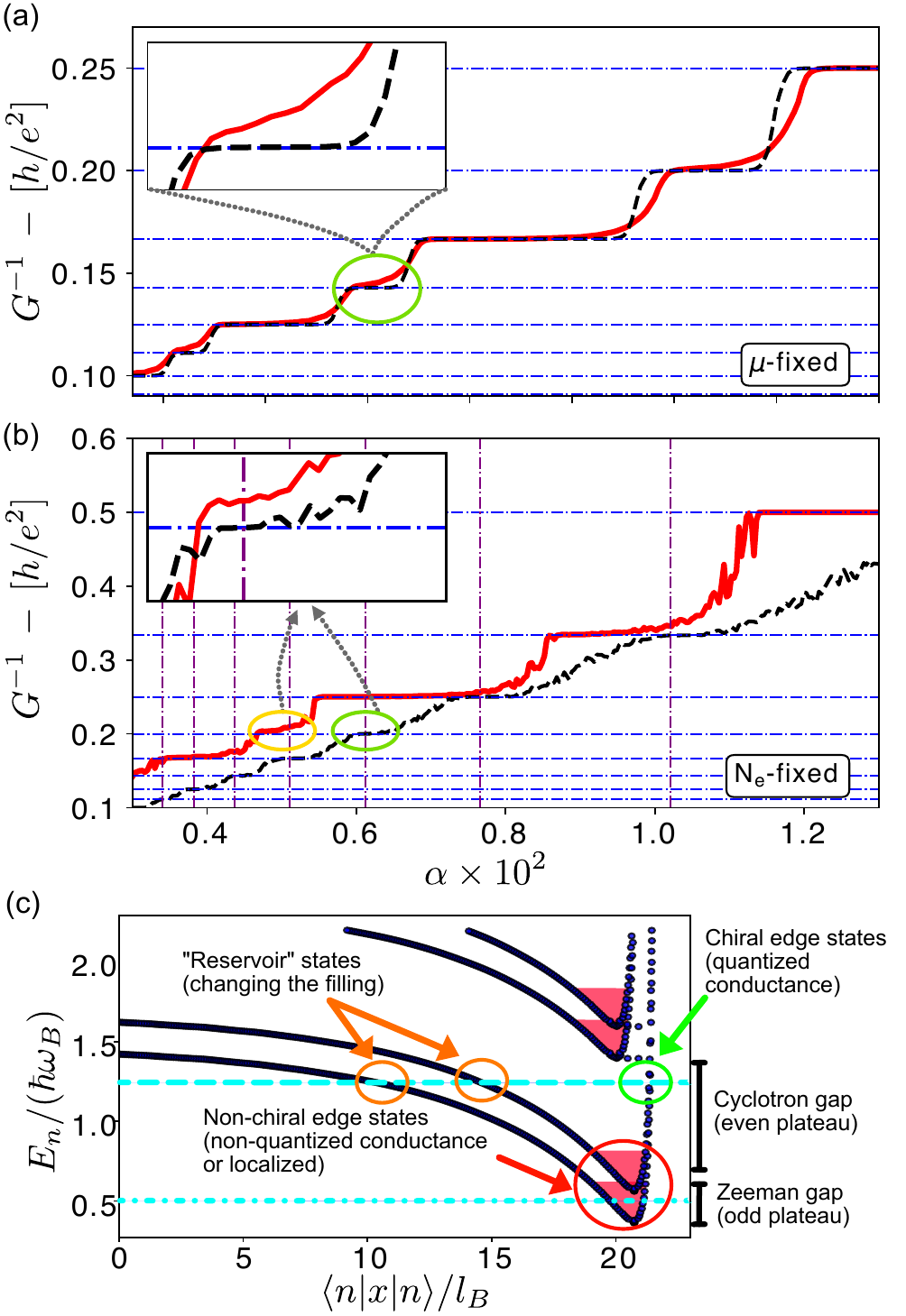}
    \caption{
    $1/G$ at finite $T$ as a function of the magnetic field parameter $\alpha$ fixing either (a) the chemical potential $\mu$ or (b) the number of particles $N_e$. The red solid and black dashed lines correspond to the same setup with and without image charges, respectively. The vertical purple dashed-dot lines highlight the integer filling factors $\nu=\alpha N_e/((N_x-3)(N_y-3))$, where $\nu=3$ is the rightmost line (here we correct for the finite size \cite{raciunas_creating_2018}). 
    (c) Sketch of the mechanism for destruction of Zeeman plateaus and elongation of cyclotron plateaus. The red shaded areas represent the states within their respective backscattering energy, the blue dashed lines represent examples of Fermi energy level, highlighting the states involved. Insets in (a) and (b) show the odd Zeeman plateaus with an estimated $\Gamma_{\ell=2}/(\hbar \omega_B) \approx 0.87$, $\Gamma_{\ell=2}/E_Z \approx 3$; for clear comparison, the red curve is shifted artificially on the black one. Other  parameters in the EM.
    }
    \label{fig:4}
\end{figure}

\noindent {\color{blue}\emph{Zeeman plateaus and temperature.}}---A further refinement of our minimal model is to introduce two additional important ingredients, namely a small finite temperature $T>0$ and the spin degree of freedom. 
For the latter, we add the term $H_Z = E_Z\sigma_Z/2$ to our single-electron Hamiltonian in Eq.~\eqref{eq:ham_1}, where $\sigma_Z$ is the Pauli-Z operator and $E_Z=  g^\star\mu_B B$ is the Zeeman energy in terms of Bohr magneton and effective gyromagnetic factor \cite{enkner_tunable_2025}. 
The conductance is then obtained by weighting the results for $T=0$ with a Fermi distribution that depends on $T$ and the chemical potential $\mu$ [see SM for more details]. 

In Fig.~\ref{fig:4} we plot the inverse conductance $1/G$ as a function of the magnetic field $B$, which we express in terms of the normalized flux per plaquette, $\alpha = e B l_0^2/(2\pi\hbar)$, in the lattice regularization ($l_0$ fixed) of the continuum LLs [see SM for details].  
In the SM we present additional simulations for more realistic parameters, which show a qualitatively similar behavior, although visually less striking due to a limited system size.

We consider two scenarios. In the first scenario shown in Fig.~\ref{fig:4}(a), we assume that the chemical potential $\mu$ is held at a fixed value for all magnetic fields, while in the plot shown in Fig. \ref{fig:4}(b), the total number of particles $N_e$ is fixed instead (see SM for details about how $N_e$ is calculated).  In both cases, we see that odd plateaus (shown in the insets) are more fragile and quantization is more easily lost when the image-charge potential is added. This can be understood from the fact that the LL separation that protects the odd plateaus is lowered by the Zeeman energy (here $E_Z\simeq \hbar\omega_B/3.5$, to mimic the values reported in \cite{appugliese_breakdown_2022,enkner_tunable_2025}, corresponding to a $g^\star$-factor of $g^\star \approx 7$. Similar simulations can be performed with any other values of $g^\star$, always leading to the same features, see \cite{de_bernardis_gihub_2025}) and becomes comparable to the strength of the backscattering threshold energy $\Gamma_0$. 
In the plot shown in Fig.~\ref{fig:3}, this would correspond to a non-quantized energy window that is as big as the corresponding plateau.  

In Fig. \ref{fig:4}(b), where $N_e$ is fixed, we observe another important feature. While the odd Zeeman plateaus are almost completely washed out by the image-charge potential, the even plateaus become wider for $U_{\rm im}\neq0$. It is known that, in general, the width of an IQH plateau increases with increasing disorder, since there are more localized states in the bulk that act as a charge reservoir. Similarly, we can ascribe the increased width of the even plateaus to the appearance of additional disorder-localized states at the bottom of the electrostatic well. 
This behaviour is very similar to voltage-defined narrow constriction, studied for example in Refs. \cite{horas_investigations_2008, siddiki_interaction-mediated_2009}.

The main mechanism for the elongation of even plateaus and destruction of the odd ones in our single-particle model is represented schematically in Fig. \ref{fig:4}(c).
Interestingly, there is no strict breakdown of the topological protection, since the cyclotron gap remains clearly open.

\noindent  {\color{blue}{\it Experimentally relevant parameters.}}---
Here, we provide an estimate of the key parameters characterizing our system, in accordance with the experimental setup reported in Ref.~\cite{appugliese_breakdown_2022}. 
The relevant parameters for electrons in a magnetic field are summarized in Tab.~\ref{tab:electrons}, while those corresponding to the split-ring resonator are listed in Tab.~\ref{tab:resonator}.

\begin{table}[h]
    \centering
     \vspace{0.2cm}
    \begin{tabular}{|l|c|}
        \hline
        $B$ & $1~{\rm T}$  \\
        $(\epsilon/\epsilon_0)$ & 13\\
        $(m/m_e)$ & 0.067\\
        $l_B$ & $25.7~{\rm nm} $  \\
        $\hbar \omega_B$ &$ 1.73~{\rm meV}$ \\
        $L_x$ & $40$~$\mu{\rm m}$\\
        $L_y$ & $200$~$\mu{\rm m}$\\
        $N_{\rm L}$ & $1.93\times 10^6$\\
        $E_Z$ & $0.2~{\rm meV}$\\
        $T$ & 50 mK \\
        \hline
    \end{tabular}
    \caption{Realistic values of the parameters for the electrons in GaAs subjected to magnetic field  \cite{appugliese_breakdown_2022}.}
    \label{tab:electrons}
\end{table}

\begin{table}[h]
    \centering
    \vspace{0.2cm}
    \begin{tabular}{|c|c|c|c|c|c|}
        \hline
        $d_{\rm edge}$ & $d$ & $\hbar\omega_{\rm LC}$ & $C$ & $\mathcal{E}_{\rm vac}$ & $\mathcal{E}_{\rm img}$ \\
        \hline
        $200~{\rm nm}$ & $40~\mu{\rm m}$ & $1~{\rm meV}$ & $100~{\rm fF}$ & $1~{\rm V/m}$ & $1~{\rm kV/m}$\\
        \hline
    \end{tabular}
    \caption{Realistic values of the parameters for the split-ring resonator \cite{appugliese_breakdown_2022}.}
    \label{tab:resonator}
\end{table}
The image electric field is given by $\mathcal{E}_{\rm img} = e/(8\pi\epsilon d_{\rm edge}^2)\approx 0.055/(d_{\rm edge}[\mu {\rm m}])^2\times 10^3\, $V/m ($d_{\rm edge}$ must be expressed in $\mu$m).

\noindent  {\color{blue}{\it Figures parameters.}}---We now summarize the microscopic parameters used to obtain the results presented in the main text.
All numerical simulations are based on the Harper–Hofstadter discretization scheme described in the Supplementary Information.

For convenience, we define the parameter
\begin{equation}
    U_B = \frac{e^2}{4\pi\epsilon l_B} = 4.3\sqrt{B[\text{T}]}~\text{meV},
\end{equation}

which quantifies the strength of the image-charge electric field. Expressing the results in terms of dimensionless ratios such as $U_B/(\hbar\omega_B)$ or $U_B/E_Z$ enables straightforward rescaling for different magnetic-field values.

For Fig. \ref{fig:2} we have $\alpha=1/80$, $N_x=N_y=70$, $Al_B^6/(\hbar\omega_B) =2\times 10^{-5}$, $\Gamma/(\hbar\omega_B)=0.4$, $J=1$.

For Fig. \ref{fig:3} parameters are $N_x=500, N_y=500$, $\alpha=0.01$, $\omega_B/J = 0.12$, $l_B/l_0\approx 4$, $U_B/(\hbar \omega_B)=80$, $E_{\rm dis}/(\hbar \omega_B) = 0.01$, $\xi_c/l_B = 1.25$. All the simulations are averaged over 10 disorder realizations.

For Fig. \ref{fig:4}(a) parameters are $N_x=120$, $N_y=80$, $\alpha_0=0.005$, $l_{B_0}/l_0\approx 5.6$, $\omega_{B_0}/J=0.062$, $\xi_c = l_{B_0}/2$, $E_{\rm dis}/(\hbar \omega_{B_0}) = 0.025$, $U_{B_0}/(\hbar \omega_{B_0}) = 50$, $d_{\rm edge}/l_{B_0}=10$, $E_Z=\hbar\omega_{B}/3.5$, $k_bT/(\hbar \omega_{B_0})=1/50$, $\mu/(\hbar \omega_{B_0})= 5.5$.
The subscript `$0$' indicates the lowest value of $\alpha$ from which the sweep starts (e.g. the lowest value of the dimensionless magnetic flux is $\alpha_0$).
The subscript `$0$' denotes the lowest value of $\alpha$ from which the magnetic-field sweep starts (i.e., the minimum value of the dimensionless magnetic flux is $\alpha_0$).
At $B = 1$ T, the corresponding value of $U_{B_0}$ is $e^2/(8\pi\epsilon) \approx 2.16$ meV $\mu$m, approximately forty times larger than the value obtained using realistic GaAs parameters.
The chemical potential in is fixed differently for red and black lines, such that the minimum energy correspond to 0 in both cases. 

For Fig. \ref{fig:4}(b) parameters are $N_x=200$, $N_y=200$, $\alpha_0=0.003$, $l_{B_0}/l_0\approx 7.3$, $\omega_{B_0}/J=0.038$, $\xi_c = l_{B_0}/2$, $E_{\rm dis}/(\hbar \omega_{B_0}) = 0.03$, $U_{B_0}/(\hbar \omega_{B_0}) = 100$, $d_{\rm edge}/l_{B_0}=10$, $E_Z=\hbar\omega_{B}/3$, $k_bT/(\hbar \omega_{B_0})=1/80$.
The number of particles in (b) is fixed at $N_e=1188$ (237) for the case with (without) image charges, such that the chemical potential is approximately the same in both cases.
At $B = 1$ T, this choice corresponds to $e^2/(8\pi\epsilon) \approx 4.33$ meV $\mu$m, roughly eighty times larger than the realistic estimate for GaAs-based systems. These enhanced values are used solely to magnify the effect for numerical analysis.
Assuming $l_B\approx 25\,$nm, the densities used relative to $N_e$s reported above are $n_e\approx 2.6\times 10^{-3}\,$nm$^{-2}$ $=2.6 \times 10^{11}\,$cm$^{-2}$ ($0.5\times 10^{-3}\,$nm$^{-2}$ $=0.5 \times 10^{11}\,$cm$^{-2}$).

\clearpage

\setcounter{section}{0}
\setcounter{subsection}{0}
\setcounter{equation}{0}%
\setcounter{figure}{0}%
\setcounter{table}{0}%

\setcounter{page}{1}

\renewcommand{\thesection}{S\arabic{section}}
\renewcommand{\thesubsection}{S\arabic{section}.\arabic{subsection}}
\renewcommand{\theequation}{S\arabic{equation}}
\renewcommand{\thefigure}{S\arabic{figure}}
\renewcommand{\thetable}{S\arabic{table}}

\onecolumngrid

\begin{center}
\textbf{\Large Supplemental Material for
\\``Electrostatics-induced breakdown of the integer quantum Hall effect in cavity QED''}
\bigskip

Gian Marcello Andolina,$^{1}$
Zeno Bacciconi,$^{2}$
Alberto Nardin,$^{3}$
Marco Schirò,$^{1}$
Peter Rabl,$^{4,5,6}$
and Daniele De Bernardis$^{7}$

\bigskip

$^{1}${\it JEIP, UAR 3573 CNRS, Collège de France, PSL Research University, 11 Place Marcelin Berthelot, F-75321 Paris, France}

$^{2}${\it SISSA-International School for Advances Studies, via Bonomea 265, 34136 Trieste, Italy
ICTP-The Abdus Salam International Centre for Theoretical Physics,
Strada Costiera 11, 34151 Trieste, Italy}

$^{3}${\it Universite Paris-Saclay, CNRS, LPTMS, 91405, Orsay, France}

$^{4}${\it Walther-Meißner-Institut, Bayerische Akademie der Wissenschaften, 85748 Garching, Germany}

$^{5}${\it Technische Universität München, TUM School of Natural Sciences, Physics Department, 85748 Garching, Germany}

$^{6}${\it Munich Center for Quantum Science and Technology (MCQST), 80799 Munich, Germany}

$^{7}${\it National Institute of Optics [Consiglio Nazionale delle Ricerche CNR–INO], care of European Laboratory for Non-Linear Spectroscopy (LENS), Via Nello Carrara 1, Sesto Fiorentino, 50019, Italy}

\bigskip

In this Supplemental Material, we provide additional information on the theoretical model used and the numerical methods used to solve it. 
Additionally, we report a detailed analysis of the non-quantized conductance and the effect of the image charges compared to the effect of cavity-vacuum corrections.
\end{center}

\maketitle
\tableofcontents
\newpage

\twocolumngrid

\section{Details on electrons in a uniform magnetic field}\label{sec:landau_problem}
In the absence of any external potential,  $U(\mathbf{r}) = 0$, the single-electron Hamiltonian in Eq.~\eqref{eq:ham_1} reduces to
\begin{equation} \label{eq:H_magnetic}
{H}_e = \frac{1}{2m} \left[ p_x^2 + \left( p_y + eB~x \right)^2 \right]~.
\end{equation}
This describes a charged particle in a uniform magnetic field $B$, taken in the Landau gauge $\mathbf{A}_{\rm ext} = B(0,x)$. The eigenfunctions of ${H}_e$ are denoted as $\ket{\ell, k}$. The wavefunctions $\psi_{\ell,k}(\mathbf{r})\equiv\langle{{\bf r}|\ell,k}\rangle$ read
\begin{equation} \label{eq:psi_magnetic}
\psi_{\ell,k}(\mathbf{r}) = \frac{1}{\sqrt{L_y}}~e^{i k y}~ \phi_{\ell}(x - x_k)~,
\end{equation}
where $\phi_{\ell}(x - x_k)$ are harmonic oscillator eigenfunctions centered at the guiding center position $x_k = - kl_B^2$, where $l_B=(\hbar/eB)^{1/2}$. The corresponding energy spectrum consists of macroscopically degenerate Landau levels, $E_\ell = \hbar \omega_B \left( \ell + \frac{1}{2} \right)$ and $\omega_B=eB/m$. The degeneracy arises from the translational invariance along the $y$-direction: for each allowed wavevector $k$, the eigenstate is localized along $x = x_k$, but carries a plane wave in $y$. In a finite system of length $L_y$, the quantization of $k = (2\pi n / L_y)$ leads to a number of available guiding centers proportional to the magnetic flux through the sample, yielding a degeneracy per Landau level of 
\begin{equation} \label{eq:degeneracy}
N_{\rm L} = \frac{L_xL_y}{ 2\pi l_B^2}~,
\end{equation}
where $L_xL_y$ is the sample area.

This picture can be extended to the case of a slowly varying potential $U(x)$.  Linearizing the potential around the position $x_i$,  for sufficiently small $\delta x_i = x-x_i$, we write
\begin{equation}\label{eq:Linear}
U(x) \approx U(x_i) + \partial_x U(x)|_{x=x_i}~ \delta x_i~,
\end{equation}
and substitute into the single-particle Hamiltonian Eq~\eqref{eq:H_magnetic}:
\begin{equation} \label{eq:H_magnetic1}
{H}_e = \frac{1}{2m} \left[ p_x^2 + \left( p_y + eB~x \right)^2 \right] + U(x_i)+U'(x_i)(x-x_i)~.
\end{equation}
This describes a harmonic oscillator in the $x$-direction with cyclotron frequency $\omega_B$, subject to a uniform electrostatic force $F = -\partial_x U(x)|_{x=x_i}$. The effect of the potential gradient is to shift the oscillator center.

The eigenfunctions retain the form of plane waves along $y$, with a displaced guiding center:
\begin{equation}
\psi_{\ell,k}(\mathbf{r}) = \frac{1}{\sqrt{L_y}}~ e^{i k y}~ \phi_\ell(x - x_k^i)~,
\end{equation}
where the shifted guiding center is given by
\begin{equation}
x_k^i = - k \ell_B^2 - \frac{\partial_x U(x)|_{x=x_i}}{m \omega_B^2}~.
\end{equation}
The linear approximation (Eq.~\eqref{eq:Linear}) shoud thus applied to the case where $x_i=x_k^i$. Hence, to leading order in the gradient expansion, the Landau level energies become
\begin{equation} \label{eq:energy_inhomogeneous}
E_{\ell,k}(x_i) \approx \hbar \omega_B \left( \ell + \frac{1}{2} \right) + U\left(-k \ell_B^2 - \frac{\partial_x U(x)|_{x=x_i}}{m \omega_B^2}\right)~.
\end{equation}
Thus, the Landau levels are rigidly shifted by the local potential $U(x_i)$, preserving the degeneracy associated with translation in $y$, while the real-space localization of the wavefunctions is modulated via the position $x_i$. 
By means of this equation, we can calculate the velocity along $y$ in the position $x_i$ as $v_y(x_i)= \frac{1}{\hbar} \partial_k E_{\ell,k}$. Hence, the $y-$ velocity reads
\begin{equation} \label{eq:v}
v_y(x_i)=-\frac{l_B^2}{\hbar} \partial_x U(x)|_{x=x_i}~.
\end{equation}
This equation establishes that the $y$-velocity is proportional to the gradient of the potential, $\partial_x U(x)|_{x=x_i}$.

\section{Numerical methods}
\label{sec:numerical_methods}
In this section we describe in detail the numerical methods used to obtain all the figures in this work.

\subsection*{Real-space numerical discretization: the Harper-Hofstadter model}
\label{app:harper_hof}
The single-particle solutions of Eq.~\eqref{eq:ham_1} have been numerically obtained by solving a discrete model,
the Harper–Hofstadter Hamiltonian~\cite{de_bernardis_light-matter_2021,de_bernardis_chiral_2023}, 
which has Eq.~\eqref{eq:ham_1} as its continuum limit when the magnetic flux per square plaquette is small, as we are now going to briefly review. 
On a square lattice of dimensions $N_x \times N_y$, the aforementioned Hamiltonian reads:
\begin{equation}\label{eq:ham_HH}
\begin{split}
    H_{\rm HH} &= \sum_{i}\Big[4\hbar J + U(\textbf{r}_i)\Big]{c}_i^{\dagger}{c}_i-\hbar J \sum_{\langle{i,j\rangle}}\left[e^{i\phi_{ij}}{c}_i^{\dagger}{c}_j + {\rm H.c.} \right].
\end{split}
\end{equation}
Here, ${c}_i$ (${c}_i^\dagger$) destroys (creates) an electron at lattice position $\mathbf{r}_i$, $J$ is the hopping frequency, and the sum $\sum_{\langle{i,j\rangle}}$ runs over nearest-neighbor lattice sites. In the Landau gauge, the magnetic phase factor is given by
\begin{equation}\label{eq:phase}
\begin{split}
    \phi_{ij} = -\frac{\pi \alpha (y_i - y_j)(x_i + x_j)}{l_0^2}~.
\end{split}
\end{equation}
where $l_0$ is the lattice spacing and $\alpha = e B l_0^2/(2\pi\hbar)$ is the magnetic flux through each lattice plaquette. In the continuum limit, obtained by taking $\alpha \to 0$ and $J \to \infty$ while keeping their product fixed, the Harper–Hofstadter model reduces exactly to the standard model of a single particle in a uniform magnetic field. In this limit, the relevant physical parameters can be expressed in terms of the simulation parameters $\alpha$ and $J$ as follows:
\begin{equation}\label{eq:Parameters}
\begin{split}
 &m=\frac{\hbar}{2Jl_0^2}~,\\
& \omega_{B}= 4\pi\alpha J ~,\\
&l_B=\frac{l_0}{\sqrt{2\pi\alpha}}~.
\end{split}
\end{equation}
Leading-order corrections to energies due to lattice discretization scale as $(l_0/l_B)^2=2
\pi\alpha$ \cite{de_bernardis_light-matter_2021}. The parameter regime of interest for our simulations involves large lattice sizes $N_x, N_y \sim 10^1-10^2$, small magnetic flux $\alpha \sim 10^{-1}-10^{-2}$, and hopping frequency $J$ setting the overall energy scale. Under these conditions, the continuum limit is already accurately reproduced \cite{de_bernardis_light-matter_2021}.

\subsection*{Discretization in the Landau level basis}    
\label{app:Landau_continuous}

As an alternative numerical method, we can express the Hamiltonian ${H}_e$ in the main text projected on the basis of continuous Landau levels defined in Eq. \eqref{eq:psi_magnetic}.
In this way we have
\begin{equation}
\langle{\ell, k | {H}_{e} | \ell^{\prime}, k^{\prime}\rangle} = \hbar \omega_B\left(\frac{1}{2}+\ell \right) \delta_{\ell \ell^{\prime}} + \langle{\ell, k | U(\textbf{r}) | \ell^{\prime}, k^{\prime}\rangle}.
\end{equation}
This matrix is then numerically diagonalized by truncating the Hilbert space. 
By introducing a maximum number of Landau levels $N_{{\rm max}\,\ell}$ and a maximum number of $k$-states within each Landau level $N_{{\rm max}\,k}$. Typically we take $N_{{\rm max}\,\ell}\approx 2-5$ and $N_{{\rm max}\,k}\approx 100-400$.

\subsection*{Transport simulation}

To investigate the quantized Hall conductivity, we simulate electronic transport in a two-terminal geometry, as described in the main text. We employ the Python package \texttt{Kwant}~\cite{groth_kwant_2014} to attach two leads to a finite square-lattice system whose Hamiltonian is given by Eq.~\eqref{eq:ham_HH}.

The conductance $G(E_F)$ is evaluated as a function of the Fermi energy $E_F$ in the leads using the standard scattering approach~\cite{arwas_quantum_2023}, which computes the transmission matrix from lead 1 to lead 2 for an incident state at energy $E = E_F$. The numerical implementation and data are publicly available in Ref.~\cite{de_bernardis_gihub_2025}. Further details on the transport formalism and numerical procedure can be found in Refs.~\cite{groth_kwant_2014,arwas_quantum_2023,yi-thomas_integer_2025}.

\subsection*{Finite temperature and number of particles}

Given a zero-temperature conductance profile $G(E)$, the finite-temperature conductance $G(\mu,T)$ is obtained following Ref.~\cite{yi-thomas_integer_2025} as 

\begin{equation}
    G({\mu},T) = {-}\int_{{-\infty}}^{\infty} dE' G(E') \frac{\partial}{\partial E^\prime} f(E^\prime-{\mu}, T)~,
\end{equation}
where $f(E,T)=1/(\exp\left[ E/(k_{B} T) \right] + 1)$ is the Fermi distribution, with $k_B$ the Boltzmann constant.
To maintain a fixed number of electrons, we compute the density of states $\rho_e(E)$ using the dedicated \texttt{Kwant} routine~\cite{groth_kwant_2014}. The particle number is then obtained as

\begin{equation}
N_e(\mu) = \int_{-\infty}^{\infty} dE~ f(E-\mu,T)~\rho_e(E)~.
\end{equation}

 Enforcing particle-number conservation yields a floating chemical potential $\mu = \mu(T,B)$, which depends on both temperature and magnetic field.

\section{Ramsauer--Townsend-like conductance}
\label{app:ramsauer}

In this section, we analyze in detail the origin of the quantization breakdown exposed in Fig. \ref{fig:3} of the main text.
First, we provide an analytic description using the guiding-centers approximation. In the disorderless case, the problem remaps to a single-particle scattering problem over a potential well, exhibiting the typical Ramsauer--Townsend scattering resonances.
In the fully disordered case, the scattering resonances are gone and the conductance can be obtained in an exponential form using a WKB approximation.
Moreover, we validate this analytic picture with extensive numerical simulations.

\subsection*{Analytic description of non-quantized conductance through guiding-centers}

The presence of non-chiral stationary states, such as those shown in Fig.~\ref{fig:2} of the main text, renders the conductance problem analogous to the standard scattering scenario involving bound or quasi-bound states in a potential well, as discussed below.
As we reviewed in Sec.~\ref{sec:landau_problem}, the orbits of electrons moving according to Eq.~\eqref{eq:ham_1} in the absence of any external potential are just harmonic oscillator wavefunctions centered at a position $x_k=-k l_B^2$ (cf. Eq.~\eqref{eq:psi_magnetic}).
First of all, suppose now we are concerned with the quantum-mechanical motion of a single electron whose orbit is close to the minimum of $V(x)=U_{\rm im}(x)+U_{\rm conf}(x)$ at $x=x_0$. In this case, we can expand this potential at the second order
\begin{subequations}
\label{eq:approximations_ramsauer}
\begin{equation}
    \label{eq:approximations_ramsauer_1}
    V(x) \approx V(x_0) + \frac{V''(x_0)}{2}(x-x_0)^2.
\end{equation}
Secondly, the importance of the {\it weak} disorder in our problem lies in the fact that it can introduce scattering among different $k$ states in Eq.~\eqref{eq:psi_magnetic}, since it breaks momentum conservation along the $y$ direction. 
As a simplifying assumption, retaining this essential feature but simplifying the problem considerably, we approximate
\begin{equation}
    U_{\rm dis}(x,y) \approx U_{\rm dis}(y).
\end{equation}
\end{subequations}
Namely, for this theoretical analysis, we approximate the disorder to be uniform along $x$. 

It is useful to introduce $\pi_\alpha = p_\alpha + e A_{\rm ext}(\mathbf{r})$, known as the kinetic momenta ($\alpha=x,y$). These two operators are canonically conjugate, $[\pi_\alpha,\pi_\beta]=i \hbar^2/l_B^2 \epsilon_{\alpha\beta}$. Here $\epsilon_{\alpha\beta}$ is the Levi-Civita symbol. 
A second set of operators can be introduced: $R_\alpha = r_\alpha + \frac{l_B^2}{\hbar}\epsilon_{\alpha\beta} \pi_\beta$, known as the guiding-centers.
It is straightforward to check that $[R_\alpha,R_\beta]=-i l_B^2$ and that $[\pi_\alpha,R_\beta]=0$. Namely, the guiding-centers define a second, independent set of canonically conjugate variables.

If the magnetic field is strong enough, the kinetic momenta freeze and the Landau Hamiltonian $\boldsymbol{\pi}^2/2m = \hbar\omega_B(\ell+1/2)$ becomes just the usual constant Landau-level cyclotron energy; on the other hand, at large magnetic field, the electron coordinates $r_\alpha \simeq R_\alpha$ can be replaced by their guiding centers. 
In this limit, and with the approximations Eq.~\eqref{eq:approximations_ramsauer},  Eq.~\eqref{eq:ham_1} becomes
\begin{equation}
    H_e \approx \widetilde V_0 + \frac{V''(x_0)}{2}R_x^2 + U_{\rm dis}(R_y),
\end{equation}
where $\widetilde V_0 = \hbar\omega_B\left(\ell+\frac 12\right) + V_0$.
Notice that, since $R_x$ and $R_y$ do not commute, this is a genuine 1D quantum problem. 
It is instructive to inspect Heisenberg's equation of motion of $R_x(t)$ and $R_y(t)$:
\begin{subequations}
    \begin{equation}
        \dot R_y(t) = \frac{l_B^2}{\hbar} V''(x_0) R_x
    \end{equation}
    and
    \begin{equation}
        \dot R_x(t) = -\frac{l_B^2}{\hbar} U_{\rm dis}'(R_y).
    \end{equation}
\end{subequations}
In the absence of disorder, the second equation tells us that $\langle R_x(t)\rangle = \langle R_x(0)\rangle $ is constant in time; the first equation then gives $\langle R_y(t) \rangle=\langle R_y(0) \rangle + \frac{l_B^2}{\hbar} V''(x_0) \langle R_x(0) \rangle$: the guiding center follows the equipotential without backscattering, the direction being set by $\langle R_x(0) \rangle$.
On the other hand, if disorder is present $\langle R_x\rangle$ can change and the electron can reverse its direction, effectively backscattering.

Since the commutator $[R_x,R_y] =-i l_B^2$ exactly mimics the canonical momentum-position commutation relations $[p,x]=-i\hbar$, we now notice that the problem is {\rm equivalent} to a 1D Schr\"odinger equation.
In particular, writing $R_x=-i l_B^2 \frac{\partial}{\partial R_y}$ as the canonical momentum associated to the position $R_y$, we get
\begin{equation}
    \label{eq:app_effective_schrodinger}
    H_e \approx - \frac{\hbar^2}{2M_{\rm eff}} \frac{\partial^2}{\partial R_y^2} + \widetilde U_{\rm dis}(R_y)
\end{equation}
where we identified an effective mass as $M_{\rm eff}^{-1}= V''(x_0) l_B^4/\hbar^2$ and we shifted $\widetilde U_{\rm dis} = U_{\rm dis}+\widetilde V_0$.

When an electron is injected at one side of the sample at a Fermi energy $E_F$ which is close to the bottom of the image-potential pocket, yet above the disorder $U_{\rm dis}$, it has a small quantum-mechanical above-barrier reflection probability.
For a long system with smoothly varying disorder, we expect a semiclassical approximation (WKB) to describe such a reflection process correctly; this yields~\cite{landau_qm}
\begin{subequations}
    \label{eq:wkb}
    \begin{equation}
        R(E_F) \sim \exp\bigl(-4\, {\rm Im}\!\left[\varphi(E_F)\right]\bigr)
    \end{equation}
    where
    \begin{equation}
        \varphi(E_F)=\frac{\sqrt{2M_{\rm eff}}}{\hbar}\int_{y_1}^{y_2} \!\!\sqrt{E_F- \widetilde U_{\rm dis}(y)}\,\,dy,
    \end{equation}
\end{subequations}
where $y_{1(/2)}$ are the (complex) turning points $E_F= \widetilde U_{\rm dis}(y)$~\cite{landau_qm}.
This motivates the exponential fit for the conductance in the main text (cf. Fig.~\ref{fig:3}).
Notice that, as can be expected intuitively, 
(i) reflection is enhanced(/suppressed) at low- (/high-) energies, and that 
(ii) when the image-charge induced pocket gets tighter (increasing $V''(x_0)$), the effective mass $M_{\rm eff}$ decreases and the above-barrier reflection (which is a purely quantum effect in this language) becomes more important.

Secondly, we would like to highlight that, when the semiclassical approximation fails, Ramsauer--Townsend-like oscillations
(resonant tunneling) can appear, as we are going to demonstrate below numerically.

Before that, we would like to emphasize how non perfectly transmitting edges (source and drain) can play the same role of the disorder in Eq.~\eqref{eq:app_effective_schrodinger} (i.e. cause the backscattering); their description is, however, arguably more complicated, and we will not delve into it.
Secondly, we want to stress that 
(i) the quadratic approximation Eq.~\eqref{eq:approximations_ramsauer_1} may be too crude to describe the transmission as a function of $E_F$ (higher-order terms in the expansion could be non-negligible, especially at large $E_F$), and that
(ii) in our case, the confinement potential is not a smooth function of $x$; this makes the identification of the energy scale $\sim M_{\rm eff}^{-1}$ in our case.
On the other hand, we expect the backscattering energy $\Gamma_\ell$ in Eq.~\eqref{eq:tunneling_energy} to play its role.
As a consequence, in the following analysis we adopt a less ``fundamental" viewpoint, focusing on this latter energy scale.

\subsection*{Numerical exploration of the Ramsauer--Townsend-like conductance}

\begin{figure}
    \centering
    \includegraphics[width=\columnwidth]{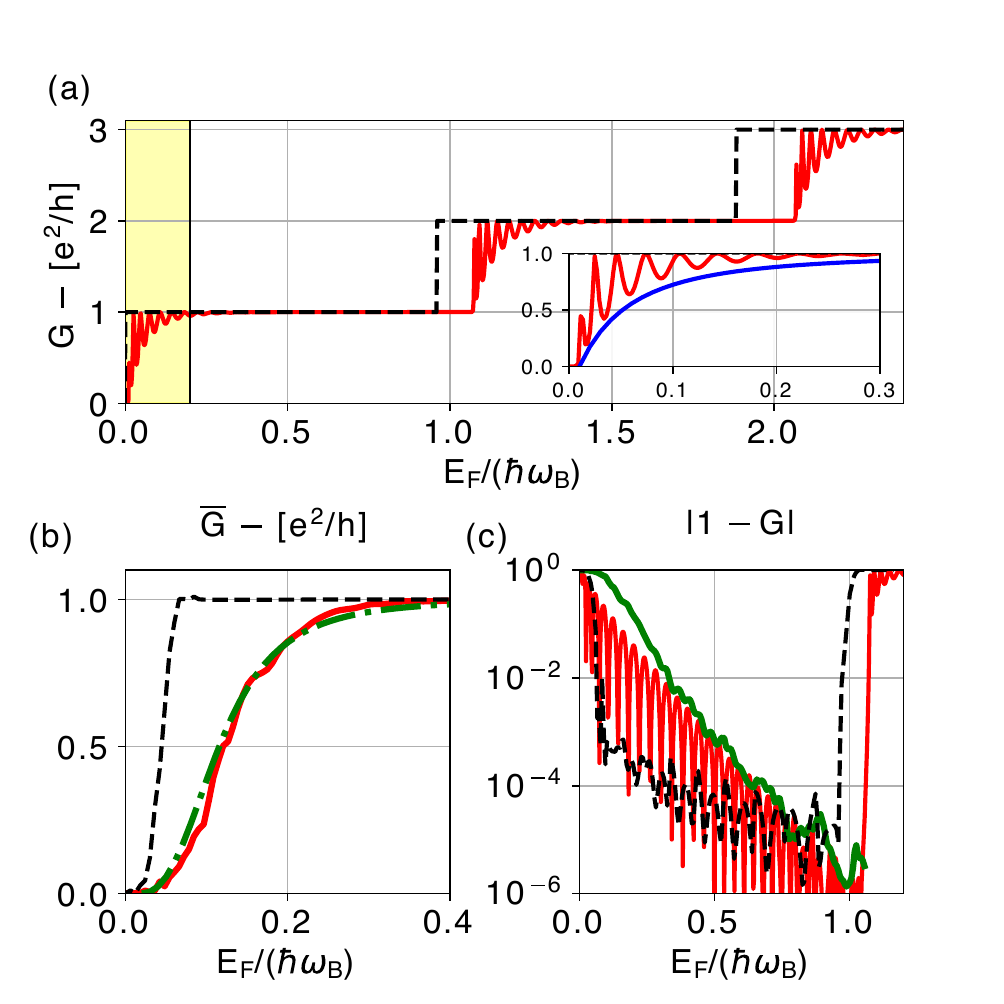}
    \caption{(a) Conductance $G(E_F)$ as a function of the Fermi energy, in units of conductance quantum $e^2/h$. The black-dashed line is the system without the image charge potential term, while the red solid line is obtained under the same parameters but also including the image charge potential in Eq. \eqref{eq:single_plate_image_pot}. The curves are shifted such that the minimum energy eigenvalue is at 0.
    The yellow shaded area is the energy range where $E_F<U_*$.
    Inset: zoom over the first $\ell=0$ plateau. The blue solid line is the analytic estimate using Eq. \eqref{eq:rams_T_approx} with $U_* = \Gamma$.
    (b) $\ell=0$ plateau with disorder, averaged over $N_{\rm dis}=50$ realizations. The green dot-dashed line is a phenomenological fit using Eq. \eqref{eq:rams_T_approx} with $U_* = \Gamma, f\approx 8, \gamma\approx 2$.
    (c) Comparison between the $\ell=0$ plateau for a clean system with images (red solid line), without images (black dashed line), and with images and disorder (green solid line), plotted in log-scale as $|1-G|$ (in units of $e^2/h$). 
    Parameters: (clean system)$N_x=200$,$N_y=50$, (disordered system)$N_x=100$, $N_y=50$, $\alpha=1/40$, $\omega_B/J \approx 0.31$, $l_B/l_0\approx 2.5$, $\Gamma/(\hbar \omega_B)=0.2$, $d_{\rm edge}/l_B=20$, $E_{\rm dis}/(\hbar \omega_B) = 0.15$ (when present), $\xi_c/l_B = 1$.}
    \label{fig:SI1}
\end{figure}

In Fig. \ref{fig:SI1}(a), we show a representative result, comparing a system without image charge potential (black dot-dashed line) and including the image charge potential.
In this case, we assume a completely clean system, where $E_{\rm dis}=0$.
As anticipated above, the presence of the image-charge pocket potential gives rise to well-visible scattering resonances (see the inset), reminiscent of the Ramsauer-Townsend effect, similarly to the scattering phenomenology through a narrow constriction \cite{beenakker_quantum_1991,szafer_PhysRevLett.62.300}. 
In this disorder-free system, 
focusing on the first plateau at $\ell=0$, we heuristically find the envelope of the oscillations in the conductance to be well described by the following sigmoid function
\begin{equation}\label{eq:rams_T_approx}
\begin{split}
    G(E_F)& \approx \frac{e^2}{h}\left[ 1+f \times A^{\gamma}(E_F) \right]^{-1},
    \\
    A(E_F) & = \frac{ U_*^2(\Gamma_0)}{4E_F(E_F+U_*(\Gamma_0))},
\end{split}
\end{equation}
with $f= 1$, $\gamma = 1,$ $U_*(\Gamma_0) = \Gamma_0$. 
Notice how this fitting function depends on the injected electron's energy only as $E_F/U_*$.
As highlighted by the yellow shaded area in Fig. \ref{fig:SI1}(a), the energy range where the deviation from quantization is the largest is set by $U_*(\Gamma_0) \approx \Gamma_0$.
The same region in the case of plateaus at larger filling $\ell>0$ can be described by $\Gamma_\ell$ as of Eq.~\eqref{eq:tunneling_energy}, accounting for the larger extension $\sim \sqrt{1+\ell} \, l_B$ of the states in the higher Landau levels \cite{page_deflection_1930}.

While important for the general understanding, the scattering resonances in the conductance can be expected to disappear in any realistic configuration like Ref. \cite{appugliese_breakdown_2022}, as evidenced by the semi-classical approximation above.
In Fig. \ref{fig:SI1}(b), we show the same simulation including strong disorder, averaged over many realizations. As expected the scattering resonances are indeed washed out. 
We fit the numerically obtained conductance with Eq. \eqref{eq:rams_T_approx}, finding $f\approx 8, \gamma\approx 2$ and 
$U_*(\Gamma_0) \approx \Gamma_0$.

In Fig. \ref{fig:SI1}(c), we make a log-scale plot of the first $\ell=0$ plateau 
for a clean (solid red line) and a disordered (solid green line) Hall bar in the presence of the image charge potential, comparing these to the conductance obtained for a disordered system {\it without} the image charge potential (black dashed line).
Even in the presence of strong disorder, without images, the conductance is quantized to good accuracy over a wider region of energies; this makes the effect of the image charge potential particularly striking.

\begin{figure}
    \centering
    \includegraphics[width=\columnwidth]{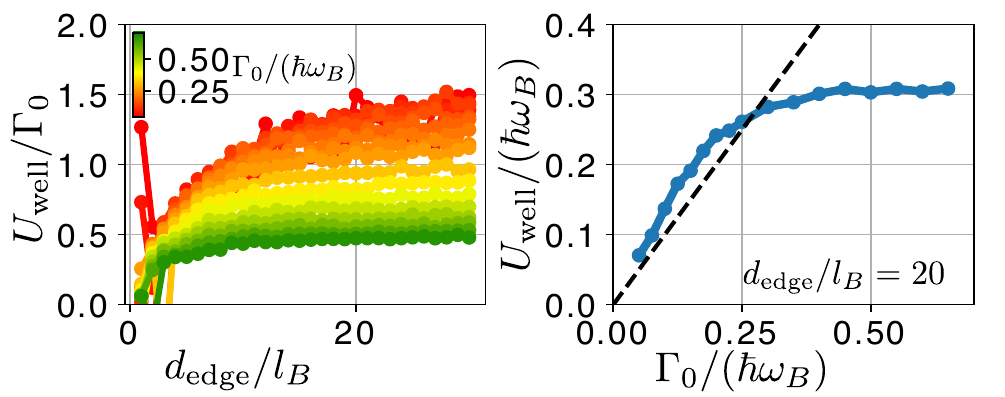}
    \caption{(Left) Effective well depth $U_*$ normalized to $\Gamma_0$, plotted
    as a function of $d_{\rm edge}/l_B$.
    (Right) $U_*$ as a function of $\Gamma_0$ keeping fixed $d_{\rm edge}/l_B$. Parameters: $N_x=N_y=50$, $\alpha=1/40$, $\omega_B/J \approx 0.31$, $l_B/l_0\approx2.5$, $\xi_c=l_B$, $E_{\rm dis}/(\hbar \omega_B)= 0.001$.}
    \label{fig:SI2}
\end{figure}

We can further numerically investigate the dependence of the fit parameter $U_*$ on the backscattering energy $\Gamma_\ell$. 
The analysis is summarized in Fig. \ref{fig:SI2}.
In the regime where $d_{\rm edge} \gg l_B$ we expect that
\begin{equation}
    U_*\approx \Gamma_0,
\end{equation}
scaling with the inverse square of the edge-plate distance $U_*\sim d_{\rm edge}^{-2}$. 
This is well visible from the data collapse in Fig. \ref{fig:SI2} left panel.
In the right panel of Fig. \ref{fig:SI2}, we instead fix $d_{\rm edge}\approx 20\,l_B$. 
We extract an approximate linear behavior $U_*\approx 4 \Gamma_0$ at small $\Gamma_0\lesssim\hbar \omega_B/4$,  
while it saturates when the backscattering energy $\Gamma_0$ approaches the Landau level spacing $\hbar\omega_B/2$. 

These data can qualitatively be interpreted within the simplified model Eq.~\eqref{eq:app_effective_schrodinger}. 
If the plate-distance $d$ is reduced, $U_*\approx\Gamma_\ell$ increases, and the transmission as described from the heuristic formula Eq.~\eqref{eq:rams_T_approx} drops.
However, as we discussed above, when the plate-distance $d$ is reduced, it can be expected that the effective mass $M_{\rm eff}\propto 1/V''(x_0)$ decreases, corresponding to a deeper and narrower image-charge-induced pocket. 
But if the effective particle becomes heavier, the above-barrier reflection becomes more important and thus the transmission through the system is suppressed according to Eq.~\eqref{eq:wkb}.
A more extensive analysis will be the subject of future work.
With these data, we can conclude that, as expected from our heuristic reasoning in the main text, the backscattering energy $\Gamma_\ell$ defines an energy-scale for the loss of quantization of the quantum Hall conductance.

\section{On the realistic value of images and backscattering energy}

Here we report additional numerical simulations in support of the discussion related to Fig. \ref{fig:3} and Fig. \ref{fig:4} in the main text.
In particular, here we explore the use of the realistic parameters discussed in the \emph{experimental parameters} section of the main text.

All the simulations are performed using \texttt{Kwant} \cite{groth_kwant_2014}, using the Harper-Hofstadter discretization framework described in Sec. \ref{app:harper_hof}. All the codes to reproduce the figures reported here are available in the repository \cite{de_bernardis_gihub_2025}.

As suggested in Ref. \cite{appugliese_cavity_2022}, the metallic structure of the cavity extends in the near proximity of the Hall bar's edge, with an approximated distance of $d_{\rm edge}\approx 200\,$nm.

Fixing the magnetic field at $B=1\,$T, accordingly to Ref. \cite{appugliese_breakdown_2022}, the filling factor is $\nu\approx 7$. 
The magnetic length is $l_B\approx 25\,$nm, and $d_{\rm edge}/l_B\approx 8$. An important energy scale is given by the Zeeman energy, which is here fixed as $E_Z = \hbar \omega_B/5\approx 0.35\,$meV (a smaller value $E_Z\approx 0.2\,$meV could also be considered \cite{enkner_tunable_2025} and is used for the experimental estimates in the main text).

With these parameters, the image charge potential is then described for all important purposes by its backscattering energy $\Gamma_\ell^{\epsilon_r}= \frac{l_B\sqrt{\ell+1}}{8\pi \epsilon_0 \epsilon_r d_{\mathrm{edge}}^2} \approx \frac{\sqrt{\ell+1}}{\epsilon_r} \,0.43 \,\mathrm{meV}$.  
As a reference, in Fig. \ref{fig:4} of the main text the relative value effective value $\Gamma_\ell^{\epsilon_r}/ E_Z$ is boosted by a factor $\sim10$ with respect to the experimental estimate, which gives $\Gamma_\ell^{\epsilon_r}/ E_Z\approx0.38$. Note that also in Fig. \ref{fig:3} we are implicitly boosting the strength of the image-charge potential. 
As commented in the text, a factor $\sim 10$ could come from different effects (sharp edges in the resonators, lower $\epsilon_r$ at the interface, terminal junctions where $d_{\mathrm{edge}}$ is effectively smaller).

To better understand the behavior at intermediate values of $\Gamma^{\epsilon_r}_\ell/E_Z$ we now effectively tune $\epsilon_r$ as $\epsilon_r^{\rm eff}$, keeping fixed the realistic value $d_{\mathrm{edge}}=200\,$nm \cite{appugliese_cavity_2022}.


\begin{figure}
    \centering
    \includegraphics[width=\columnwidth]{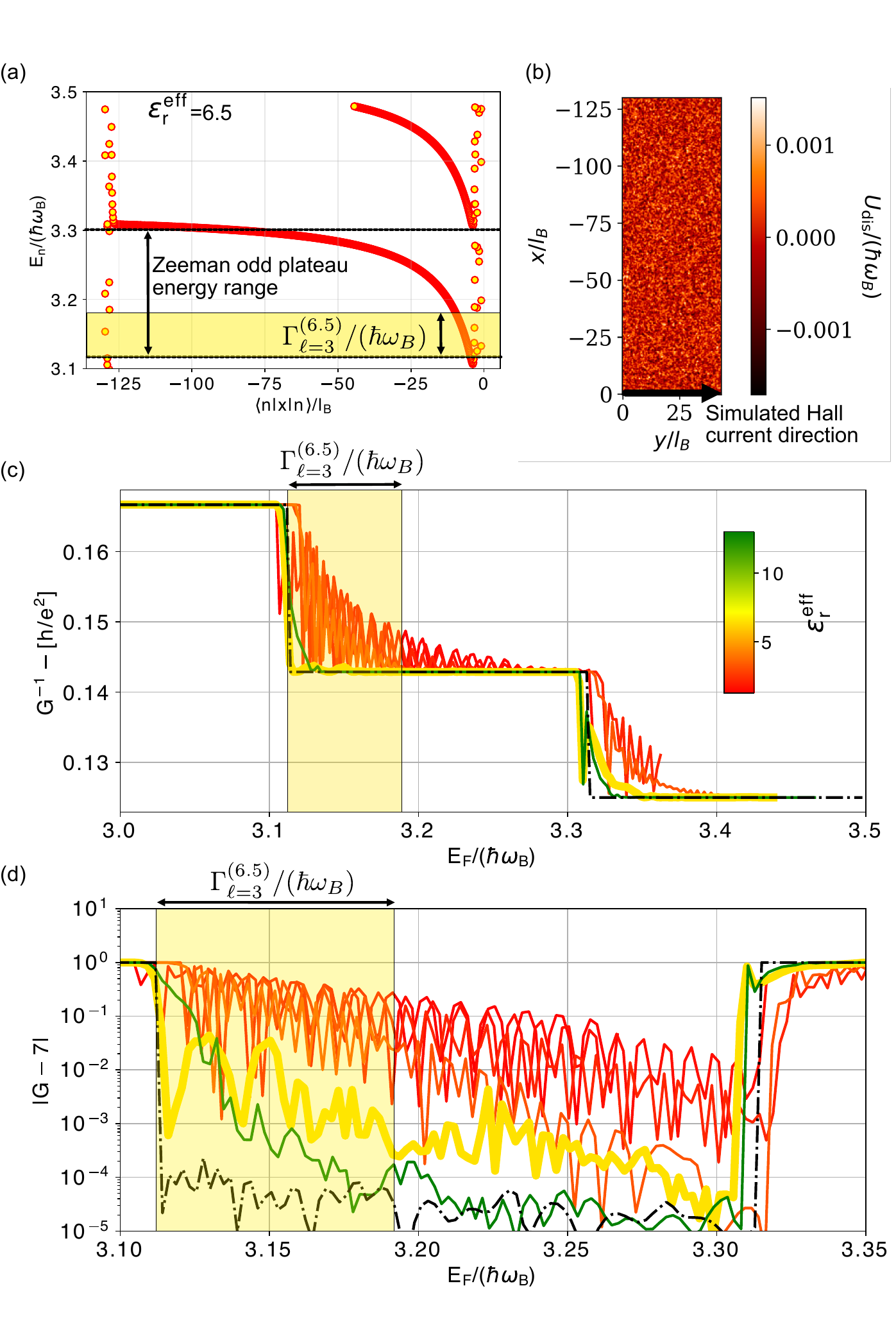}
    \caption{(a) Spectrum of the system with PBC zoomed over the eigenvalues around $\ell=3$, relevant for the simulated conductance. The red shaded area is an indicative energy range within the backscattering energy $\Gamma_{\ell =3}^{(6.5)}$ (here the number in the apex in parentheses is the respective value of the relative dielectric constant). The black dashed lines are an indicative delimitation of the energy range of the simulated plateau in the bottom panels, corresponding to the odd Zeeman plateau with $\ell=3$, and thus filling factor $\nu=7$.
    The effective relative dielectric constant here is $\epsilon_r^{\rm eff}=6.5$. (b) Disorder used in the simulation. The simulated current to compute the conductance flows on the $y$-axis, while the system is extremely elongated on $x$. Here, $x$ and $y$ axes are reversed from the usual for space issues.
    (c) The inverse conductance $G^{-1}$ as a function of the Fermi energy $E_F$ around the $\ell=3$ Landau level. The colormap represents the value of $\epsilon_r^{\rm eff}$. The yellow shaded area is the energy range of the backscattering energy of the image potential with $\epsilon_r^{\rm eff}=6.5$, $\Gamma_{\ell =3}^{(6.5)}$ (as in panel (a) the number in the apex in parentheses is the respective value of the relative dielectric constant). (d) Zoom over the $\ell=3$ odd Zeeman plateau ($\nu=7$) plotted in logscale as $|G-7|$. Same colormap as (c). The yellow shaded area is the energy range of the backscattering energy of related to the image potential of the previous panels.
    Parameters: $N_x = 300$, $N_y=100$, $\alpha=0.03$, $\omega_B/J\approx 0.38$, $l_B/l_0\approx 2.3$, $d_{\rm edge}/l_B=8$, $E_{\rm dis}=10^{-3}\hbar\omega_B$, $\xi_c=l_B/2$, $E_Z=\hbar\omega_B/5$.}
    \label{fig:SI3}
\end{figure}

In Fig. \ref{fig:SI3}, we study an example system, with low disorder, with the parameters discussed above, and a variable value for the image charges amplitude given by the effective relative dielectric constant.
Interestingly, in order to highlight the impact of both disorder and image potential, we consider a system strongly elongated along $x$, and we compute the conductance $G$ from the current flowing along the $y$-axis (as represented in Fig. \ref{fig:1} in the main text).
This choice is counterintuitive with respect to the typical representation of a two-terminal Hall bar, which is elongated in the direction of the current.
However, it is the best geometry to perform the numerical calculations since it allows one to fully appreciate the topological protection in the absence of images, minimizing the residual unquantized value of the conductance.

In Fig. \ref{fig:SI3}(a), we show the spectrum imposing periodic boundary conditions (PBC) along the $y$-axis, highlighting the presence of the pocket potential and the estimated region of backscattering states (the red shaded area). 
The simulation is performed assuming the disorder potential in Fig. \ref{fig:SI3}(b). The disorder is short-distance correlated, in agreement with the standard semiconductor literature \cite{de_bernardis_magnetic-field-induced_2022}.
Using the parameters of Fig. \ref{fig:SI3}(a), we can evaluate the backscattering energy $\Gamma_{\ell=3}^{(6.5)}/(\hbar\omega_B)\approx 0.07$, giving the depth of the pocket potential. Considering $E_Z=\hbar \omega_B/5$, we obtain $\Gamma_{\ell=3}^{(6.5)}/E_Z\approx 0.34$, predicting that $\sim 30 \%$ the conductance plateau is destroyed.
The apex $\cdot^{(6.5)}$ reports the value of the effective relative dielectric constant, that should be equal to 13 for bulk GaAs but is here chosen to be $6.5$ to have a similar estimate for $\Gamma_{\ell=4}^{(13)}/E_Z\approx 0.38$ given by the experimental parameters used in the main text. 

In Fig. \ref{fig:SI3}(c), we compute the conductance $G(E_F)$ as a function of the Fermi energy $E_F$, as in the main text, for different values of $\epsilon_r^{\rm eff}$.
The result is presented as an inverse conductance $G^{-1}$ to be comparable with the transverse resistance of the integer quantum Hall effect.
Using the bare image potential, with the GaAs screened constant, $\epsilon_r^{\rm eff}=13$, the quantization of the plateau seems to be unaltered (green solid line).
Calculating the backscattering energy, we find indeed $\Gamma_{\ell=3}^{(13)}/E_Z = 0.17$, pointing out a quite small portion of unquantized plateau.
For smaller values of the effective relative dielectric constant $\epsilon_r^{\rm eff}< 6.5$, the conductance exhibits strong oscillations, due to the Ramsauer resonances described in Sec. \ref{app:ramsauer}. 

A closer inspection in logscale in Fig. \ref{fig:SI3}(d) shows that the conductance quantization is \textit{lost} for the whole plateau extension for all the cases where $\epsilon_r^{\rm eff}\leq6.5$, as we can see from the deviation from the case without images (black dashed-dotted line).
Interestingly, the unquantized energy range seems to be larger than what is estimated through the backscattering energy, as is exemplified by the case $\epsilon_r^{\rm eff}=6.5$ (big solid yellow line, and yellow shaded area), probably due to the non-linearity of the image potential.
Also for the bare case, with $\epsilon_r^{\rm eff}= 13$ (green solid line), we observe a deviation from the quantized conductance on the $\sim 20\%$ of the whole plateau. Choosing a smaller value for the Zeeman energy, $E_Z\sim 0.2\,$meV, returns the experimental estimation in the main text, and make also the bare case not so distant from the experimental realization.

\begin{figure}
    \centering
    \includegraphics[width=\columnwidth]{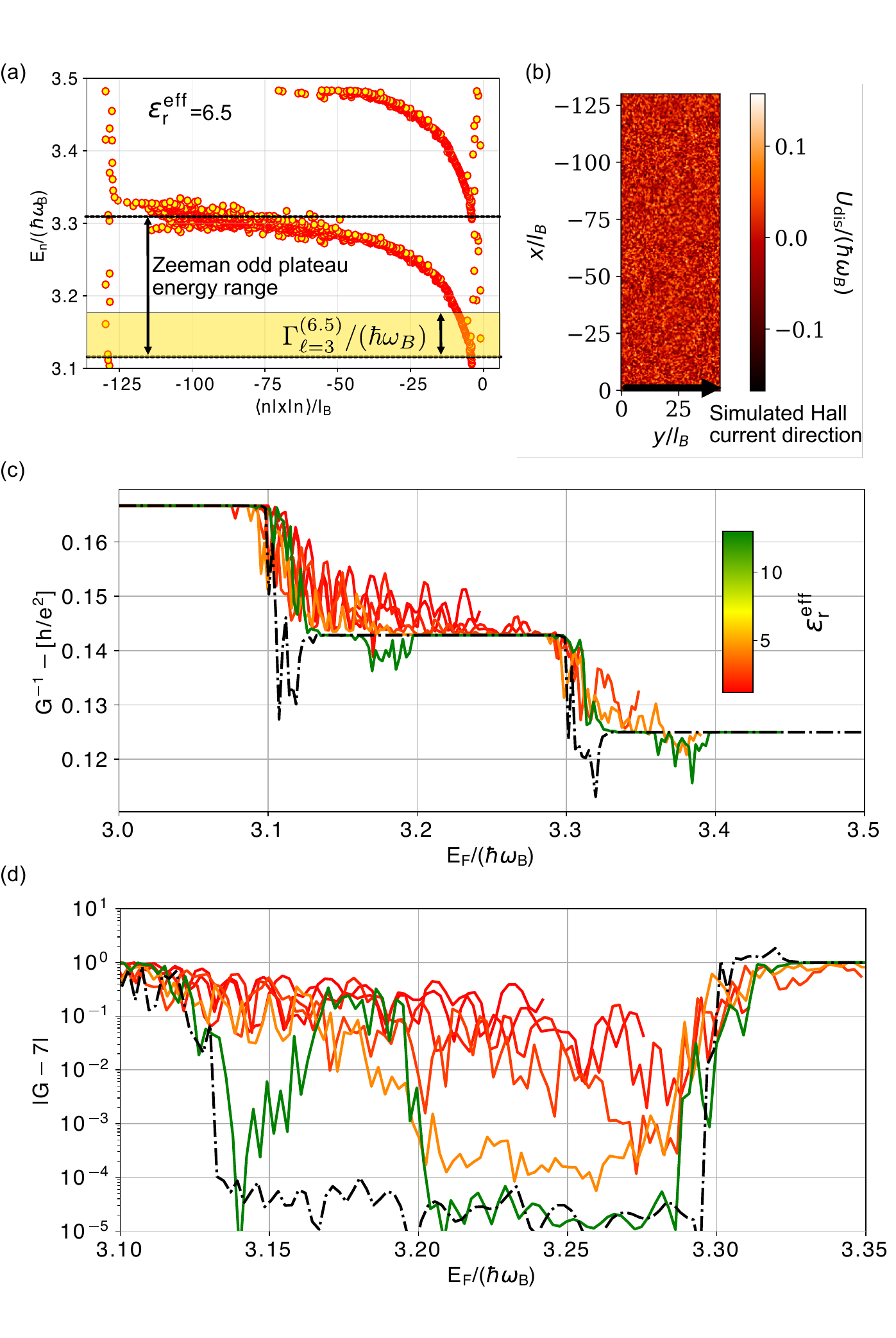}
    \caption{Same as Fig. \ref{fig:SI3}. Parameters: same as Fig. \ref{fig:SI3}, except for $E_{\rm dis}/(\hbar \omega_B) = 0.1$.}
    \label{fig:SI4}
\end{figure}

For comparison, we also include the case where the disorder is very strong, which is reported in Fig. \ref{fig:SI4}. This figure is identical to Fig. \ref{fig:SI3}, except for having the disorder boosted by a factor of $\times 10^2$.
Due to the strong broadening of the Landau level, it is not so easy to identify the energy range defined by the pocket potential, so, differently from Fig. \ref{fig:SI3}, we removed the shaded areas.
Especially from Fig. \ref{fig:SI4}(d), it is evident that the plateau without images (black dashed-dotted line) is still well quantized, despite being a bit shorted than in Fig. \ref{fig:SI3}.
On the contrary, even the bare case with images, where $\epsilon_r^{\rm eff}=13$, exhibits strong deviation from the imageless quantization.

Using the bare value for the image charge potential, following our simplified toy model, is not enough to predict a full breakdown of the plateau quantization, as the ones presented in Ref. \cite{appugliese_breakdown_2022}. However, a small enhancement $\sim 4$ possibly coming from effects beyond the toy-model analysis can be enough.


\section{Details on the dynamical split-ring resonator}

Here, we provide the derivation of the quantum mechanical Hamiltonian governing the dynamical part of the split-ring resonator. Starting from Eq.~\eqref{eq:ham_tot} in the main text, the resonator charge operator ${Q}$ can be expressed in terms of the bosonic creation and annihilation operators ${a}$ and ${a}^\dagger$ as
\begin{equation}
{Q} = \sqrt{\frac{\hbar}{2Z_0}} \left( {a} + {a}^\dagger \right)~,
\end{equation}
where $Z_0 = \sqrt{L/C}$ is the characteristic impedance of the resonator.

Substituting this expression into Eq.~\eqref{eq:ham_tot}, we obtain the full Hamiltonian in the form
\begin{equation}\label{eq:Harmonic}
{H} = {H}_e + \hbar \omega_{ LC} {a}^\dagger {a}
- e \mathcal{E}_{\rm vac} x\left( {a} + {a}^\dagger \right)+ \frac{e^2\mathcal{E}_{\rm vac}^2}{\hbar \omega_{LC}} ~ {x}^2~,
\end{equation}
where $\omega_{\rm LC} = 1/\sqrt{LC}$ is the resonance frequency of the cavity and
\begin{equation}\label{eq:E_vac}
\mathcal{E}_{\rm vac} = \sqrt{\frac{\hbar \omega_{\rm LC}}{2 C d^2}}~,
\end{equation}
denotes the amplitude of the vacuum electric field. 
The quantized cavity field is thus given by ${\mathcal{E}}_{\rm cav} = \mathcal{E}_{\rm vac} \left( {a} + {a}^\dagger \right)$, and it couples linearly to the electron displacement ${x}$. The last term in Eq.~\eqref{eq:Harmonic} represents a static harmonic potential induced by the dipole self-interaction.

It is useful to evaluate the $x$-operator on the LL states $\ket{\ell,k}$ and $\ket{\ell^\prime,k^\prime}$ of Eq.~\eqref{eq:H_magnetic}, where the Landau level index $\ell$ belongs to occupied states  $\ell\leq\ell_{\rm F} $ while $\ell^\prime$ belongs to unoccupied states $\ell^\prime> \ell_{\rm F}$, and we define $\ell_{\rm F}$ as the index of the highest occupied LL. The the $x$-operator reads
\begin{equation}\label{eq:dipole}
\langle{\ell^\prime,k^\prime|x|\ell,k}\rangle=\left(\frac{l_B}{\sqrt{2}}\right)\sqrt{\ell+1}\delta_{k^\prime,k}\delta_{\ell^\prime,\ell+1}~
\end{equation}
Thus, this x-operator connects only LLs with $\ell^\prime=\ell+1$.
The form of the linear coupling in Eq.~\eqref{eq:Harmonic} combined with Eq.~\eqref{eq:dipole} fixes the {\it single electron} light-matter coupling $g_e$ to be
\begin{equation}\label{eq:g}
\hbar g_{e} = \frac{e\mathcal{E}_{\rm vac}l_B}{\sqrt{2}}  =1.5\times 10^{-5}~{\rm meV}~.
\end{equation}
Here, the last estimate equality is derived from the parameters in Tab~\ref{tab:electrons} and Tab~\ref{tab:resonator}.
The vacuum field value in this estimate is then given by

\begin{equation}\label{eq:E_vac1}
\mathcal{E}_{\rm vac} \sim 1{\rm V/m}~,
\end{equation}
which is very similar to what was discussed in Ref. \cite{appugliese_breakdown_2022}.

Notice that it is customary to calculate this value with a more phenomenological procedure than the one employed here. First, we define the geometric compression factor as
\begin{equation}\label{eq:compression}
\eta\equiv \left(\frac{d^3}{\lambda_0^3}\right)~,
\end{equation}
where $\lambda_0=(2\pi c/\omega_{LC})$ is the light free-wavelength at the cavity frequency $\omega_{LC}$. In our system, $\eta\sim 3 \times 10^{-5}$. Thus, we can use the following phenomenological formula to estimate the cavity field as
\begin{equation}\label{eq:E_vac2}
\tilde{\mathcal{E}}_{\rm vac}\equiv \sqrt{\frac{\hbar \omega_{LC}}{\epsilon \lambda_0^3\eta}} \sim 5~{\rm V/m}~.
\end{equation}

Thus the vacuum field $\tilde{\mathcal{E}}_{\rm vac}$ obtained by using a phenomenological compression factor matches the same order of magnitude of the vacuum field ${\mathcal{E}}_{\rm vac}$ in Eq.~\eqref{eq:E_vac1}, obtained with the lumped circuit model. Note that the precise value is very hard to obtain as we are neglecting mode-shape details, fringing fields and other details of the split-ring resonator that would rather require finite-element simulations to extract details of electric field. 
For a nanocavity with $d \sim 100$ nm \cite{keller_few-electron_2017}, corresponding to a smaller electromagnetic compression factor $\eta \sim 10^{-11}$ as in Refs.~\cite{arwas_quantum_2023,borici_cavity-modified_2024}, the same parameters yield a significantly enhanced vacuum electric field,
\begin{equation}\label{eq:E_vac_nano}
\mathcal{E}^{(\rm nano)}_{\rm vac} = \sqrt{\frac{\hbar \omega{LC}}{\epsilon \lambda_0^3 \eta}} \approx 5\times10^3~\text{V/m},
\end{equation}
large enough to produce measurable cavity-induced effects.

\section{Landau polaritons in the dipole gauge}
It is worth noticing that our estimation of $g_e$ is completely in agreement with the observed behavior of Landau polaritons \cite{scalari_ultrastrong_2012, paravicini-bagliani_magneto-transport_2019} when one considers the collective transitions only, because our theory represents the equivalent dipole-gauge formulation of the original one \cite{hagenmuller_ultrastrong_2010}.
To have a direct link between our formulation and the standard theory \cite{hagenmuller_ultrastrong_2010}, we review here the Landau polariton theory consisting of collective magnetoplasmon excitation of the cyclotron transition.

For the sake of simplicity, we do not consider any electronic potential, $U({\bf r})=0$. A similar analysis holds in the case of a weak disorder. The electronic LL wavefunctions are given by the solutions of Eq.~\eqref{eq:H_magnetic}, $\ket{\ell,k}$. We introduce the bright modes 
\begin{equation}\label{eq:bright_mode}
    {b}^\dagger_{\ell^\prime,\ell}\equiv \frac{1}{\sqrt{N_{\rm L}}}\sum_{k} {c}^\dagger_{\ell^\prime,k}{c}_{\ell,k} ~.
\end{equation}

Here, ${c}^\dagger_{\ell,k}$ creates an electron in the Landau level $\ell$ with momentum $k$ along the $y$ direction, and the operators ${b}_{\ell',\ell}$ in the dilute limit obey the commutation relation \cite{hopefield_PhysRev.112.1555, hagenmuller_cavity_2012}

\begin{equation}
    [{b}_{\ell',\ell},{b}^\dagger_{n',n}] \approx \delta_{\ell,n}\delta_{\ell',n'} .
\end{equation}

We restrict the Hilbert space to the manifold formed by the Fermi sea,
\begin{equation}
    |{\rm F.S.}\rangle \equiv \prod_{\ell \leq \ell_{\rm F},k} {c}^\dagger_{\ell,k}|0\rangle ,
\end{equation}

and a single particle–hole excitation,
\begin{equation}
   |\ell,k,\ell',k'\rangle \equiv {c}^\dagger_{\ell,k}~{c}_{\ell',k'}|{\rm F.S.}\rangle , 
\end{equation}

with $\ell > \ell_{\rm F}$ and $\ell' \leq \ell_{\rm F}$.

In this manifold, the many-electron dipole operator $X=\sum_n x_n$ can be approximated, retaining only the terms corresponding to the creation of electron-hole pairs and neglecting other processes as
\begin{equation}\label{eq:x_operator}
    X\approx\sum_{\ell\leq\ell_{\rm F}}\sum_{\ell^\prime>\ell_{\rm F}}\sum_{k,k^\prime}\left(\langle{\ell^\prime,k^\prime|x|\ell,k}\rangle{c}^\dagger_{\ell^\prime,k^\prime}{c}_{\ell,k}+{\rm h.c.}\right) ~.
\end{equation}

Using Eq.~\eqref{eq:dipole}, the operator $X$ can couple only the index $\ell_{\rm F}$ with $\ell_{\rm F}+1$. Thus the many-electron dipole reads in terms of the bright mode

\begin{equation}\label{eq:x_operator1}
    X=l_{B}\sqrt{\frac{N_L(\ell_{\rm F}+1)}{2}}\left({b}_{\ell_{\rm F}+1,\ell_{\rm F}}+{b}^\dagger_{\ell_{\rm F}+1,\ell_{\rm F}}\right)~.
\end{equation}

In this manifold, the position operator $x$ can be approximated, retaining only the terms corresponding to the creation of electron-hole pairs and neglecting other processes as
\begin{equation}\label{eq:x_operator}
    x\approx\sum_{\ell\leq\ell_{\rm F}}\sum_{\ell^\prime>\ell_{\rm F}}\sum_{k,k^\prime}\left(\langle{\ell^\prime,k^\prime|x|\ell,k}\rangle{c}^\dagger_{\ell^\prime,k^\prime}{c}_{\ell,k}+{\rm h.c.}\right) ~.
\end{equation}

Using Eq.~\eqref{eq:dipole}, the operator $x$ can couple only the index $\ell_{\rm F}$ with $\ell_{\rm F}+1$. Thus the $x$-operator reads in term of the bright mode
\begin{equation}\label{eq:x_operator1}
    x=l_{B}\sqrt{\frac{N_L(\ell_{\rm F}+1)}{2}}\left({b}_{\ell_{\rm F}+1,\ell_{\rm F}}+{b}^\dagger_{\ell_{\rm F}+1,\ell_{\rm F}}\right)~.
\end{equation}
Thus, the total many-body Hamiltonian projected on the single excitation manifold can be recasted as
\begin{equation} \label{eq:Harmonic1}
\begin{split}
  &{H} = \hbar\omega_B{b}^\dagger_{\ell_{\rm F}+1,\ell_{\rm F}}{b}_{\ell_{\rm F}+1,\ell_{\rm F}} + \hbar \omega_{ LC} {a}^\dagger {a}+\\
&-\hbar g_{e}\sqrt{N_{\rm L}(\ell_{\rm F}+1)} \left({b}_{\ell_{\rm F}+1,\ell_{\rm F}}+{b}^\dagger_{\ell_{\rm F}+1,\ell_{\rm F}}\right)(  a+  a^{\dag})+\\
&+ \frac{N_{\rm L} (\ell_{\rm F}+1)\hbar^2g^2_{e}}{\hbar\omega_{\rm LC}} ~ \left({b}_{\ell_{\rm F}+1,\ell_{\rm F}}+{b}^\dagger_{\ell_{\rm F}+1,\ell_{\rm F}}\right)^2~, 
\end{split}
\end{equation}
where the coupling between the split-ring cavity mode and the bright state is controlled by the {\it vacuum Rabi frequency}
\begin{equation}\label{eq:Rabi_polariton}
  \hbar \Omega_{\ell}=\sqrt{N_L(\ell+1)}~\hbar g_{e}~.
\end{equation}

For integer filling, we have that \cite{tong_lectures_2016}
\begin{equation}
    N_{\rm L}= \frac{L_xL_y}{2\pi l_B^2}\approx 10^6.
\end{equation}
The numerical estimate is given by considering $l_B\approx 25$nm, $L_x \approx 40\mu$m, $L_y\approx 100\mu$m.
For a Fermi level corresponding to $\ell_{\rm F} = 8$, the vacuum Rabi frequency is
\begin{equation}
\hbar \Omega_{\ell_{\rm F}=8} \approx 0.05~\text{meV},
\end{equation}
in quantitative agreement with experimental observations \cite{scalari_ultrastrong_2012,keller_few-electron_2017,keller_landau_2020}.

\section{Estimations of vacuum effects}

\subsection*{Derivation of the effective Hamiltonian and estimation of the vacuum contribution}

We now include the effect of the dynamical component of the split-ring resonator on the single-electron degrees of freedom. Following Ref.~\cite{ciuti_cavity-mediated_2021}, the electronic Hamiltonian $\hat{H}_e$ is considered in the presence of weak disorder, $U{\rm d}(\mathbf{r}) \neq 0$, while both the confinement and image-charge potentials are neglected, $U_{\rm c}(\mathbf{r}) = U_{\rm im}(\mathbf{r}) = 0$. 
For a simplified estimation, we evaluate Eq.~\eqref{eq:ham_tot} within the single-electron approximation \cite{ciuti_cavity-mediated_2021,arwas_quantum_2023}. As in the rest of the letter, the contribution of the Fermi-level background is neglected.

Before performing the adiabatic elimination of higher Landau levels and the cavity degree of freedom, we switch to the Coulomb gauge. This choice facilitates direct comparison with previous works \cite{ciuti_cavity-mediated_2021,borici_cavity-modified_2024} and avoids the extensive terms that arise in the dipole gauge, which complicate the adiabatic procedure.
Applying the unitary transformation
\begin{equation}    
U=\exp\Bigl( (a-a^ \dagger\,) Q_e /2C\Bigl)~,
\end{equation}

to Eq.~\eqref{eq:Harmonic}, we obtain the Coulomb-gauge Hamiltonian:

\begin{align}\label{eq:Hcoulomb}
    &{H}^c={U}{H}{U}^\dagger = H_e \nonumber - \frac{(e\mathcal{E}_{\rm vac}l_B)^2}{\sqrt{2}}\frac{\omega_B}{\omega_{LC}} \frac{l_B}{\hbar} p_x 
i({a}-{a}^\dagger)+\\& - \frac{\omega_B}{\omega_{LC}}\frac{(e\mathcal{E}_{\rm vac}l_B)^2}{2\hbar\omega_{LC}} (a-a^\dagger)^2  +\hbar\omega_{LC} {a}^\dagger{a}~.
\end{align}

The second line of Eq.~\eqref{eq:Hcoulomb} contains the paramagnetic and diamagnetic interactions.
The first term, proportional to $p_x i(a - a^\dagger)$, represents the paramagnetic coupling between the electronic momentum and the quantized cavity field, corresponding to the linear $\bf{p}\cdot\bf{A}$ interaction. 
The second term, proportional to $(a - a^\dagger)^2$, is the diamagnetic contribution, originating from the $ \mathbf{A}^2 $ part of the minimal-coupling Hamiltonian. This term guarantees gauge invariance and stabilizes the system by counterbalancing the paramagnetic interaction, ensuring a bounded ground state and preventing unphysical superradiant instabilities.

We denote the single-particle eigenstates of ${H}_e$ in Eq.~\eqref{eq:ham_1} by $\ket{\ell,i}$, where $\ell$ is the Landau level index and $i$ labels the states split by disorder. These satisfy
\begin{equation}
{H}_e \ket{\ell, i} = E_{\ell,i} \ket{\ell, i}~,
\end{equation}
with energies
\begin{equation}
E_{\ell,i} = \hbar \omega_B \ell + \delta E_{\ell,i},
\end{equation}

where $\delta E_{\ell,i} \ll \hbar \omega_B$ characterizes the weak-disorder regime. In addition, we consider energies well below both the cyclotron and photonic scales, $E \ll \hbar\omega_B,\hbar\omega_{LC}$, so that cyclotron and cavity excitations remain far off resonance. Under these conditions, the dynamics can be projected onto the LLL and the photon vacuum, with higher electronic and photonic modes contributing only through virtual processes.

In this regime, the coupling term in Eq.~\eqref{eq:Hcoulomb} can be adiabatically eliminated through a Schrieffer–Wolff transformation, yielding an effective LLL Hamiltonian decomposed into three contributions:
\begin{equation}\label{eq:Heff_app}
H_{\rm eff} = H^{({\rm LLL})}_e + H_{\rm dia} + H_{\rm para}.
\end{equation}
The first term, $H^{({\rm LLL})}_e \equiv \mathcal{P} H_e \mathcal{P}$, is the electronic Hamiltonian projected onto the LLL, where the operator
\begin{equation}
    \mathcal{P} = \sum_i \ket{0,i}\bra{0,i} \otimes \ket{0_{\rm ph}}\bra{0_{\rm ph}}
\end{equation}

projects onto both the LLL and the photon vacuum.
The second term is the diamagnetic contribution, which—being already quadratic in $g_e$—is simply projected onto the same subspace:
\begin{equation}\label{eq:Ha2}
H_{\rm dia} = \frac{\omega_B}{\omega_{LC}} \frac{(e\mathcal{E}_{\rm vac}l_B)^2}{2\hbar \omega_{LC}}
\mathcal{P} (a - a^\dagger)^2 \mathcal{P}.
\end{equation}

Finally, we use second-order perturbation theory  \cite{andolina_amperean_2024} to obtain the effect of the vacuum paramagnetic term
\begin{equation}\label{eq:Hpa}
    {H}_{\rm para} = -\frac{(e\mathcal{E}_{\rm vac}l_B)^2}{2\hbar\omega_{LC}}\frac{\omega_B}{\omega_{LC}} \mathcal{K},
\end{equation}

where 
\begin{equation}
\label{eq:K-kernel}
\begin{split}
     \mathcal{K}= & \frac{1}{2}\sum_{ \ell,j}|0,i\rangle {\langle{ 0,i|\frac{l_B p_x}{\hbar}|\ell,j \rangle}\langle{ \ell,j|\frac{l_B p_x}{\hbar}|0,i^\prime \rangle} } \times
        \\
        &\times  \left( \frac{\omega_B}{\omega_{LC}+\Delta_{\ell,i,j} } + \frac{\omega_B}{\omega_{LC}+\Delta_{\ell,i^{\prime},j} } \right) \langle 0,i^\prime |~,
\end{split}
\end{equation}

and $\hbar \Delta_{\ell,i,j}$ denotes the electronic transition energy,
$\hbar \Delta_{\ell,i,j} = E_{\ell,j} - E_{0,i}$. Since the diamagnetic gives a rigid shift, the paramagnetic term is the most important contribution stemming from the vacuum field. We now proceed to give an estimate of that.

To estimate the magnitude of Eq.~\eqref{eq:Hpa}, we note that the contribution of $\mathcal{K}$ is of order one, as will be shown momentarily. Therefore, we can evaluate the dimensional prefactor of Eq.~\eqref{eq:Hpa} by considering $\mathcal{E}_{\rm vac}=1~{\rm V/m}$, $\hbar \omega_{LC}=1$ meV and $l_B = 25~{\rm nm}$, as reported in Tables~\ref{tab:electrons}–\ref{tab:resonator}.
\begin{equation}\label{eq:Heff_E_scale}
\frac{(e\mathcal{E}_{\rm vac}l_B)^2}{2\hbar\omega_{LC}}\frac{\omega_B}{\omega_{LC}}
\approx 3\times10^{-10}~{\rm meV} \frac{\omega_B}{\omega_{LC}}.
\end{equation}
Even assuming $\omega_B / \omega_{LC} \approx10^2$, the resulting energy scale remains on the order of $10^{-8}$~meV. Hence, the prefactor is small, leading to a strong suppression of vacuum-induced effects at experimentally relevant energy scales and for realistic parameter regimes. This enstimation is quite sentive to compression factor: using the vacuum field of nanocavities, $\mathcal{E}^{(\rm nano)}_{\rm vac}\approx1~{\rm kV/m}$ (see Eq.~\eqref{eq:E_vac_nano}), instead of $\mathcal{E}_{\rm vac}\approx~{\rm V/m}$ as in Refs.~\cite{arwas_quantum_2023, borici_cavity-modified_2024}, enhances the vacuum prefactor by six orders of magnitude, yielding a value of the order of $10^{-2}$~meV, which can in turn lead to the breakdown of the quantized conductance.

We now examine the operator $\mathcal{K}$ in Eq.~\eqref{eq:K-kernel}. This dimensionless term includes a sum over the Landau-level degeneracy $N_{\rm L}$. Given that $N_{\rm L} \approx 10^6$, one might expect such a large degeneracy to strongly enhance the effect, potentially up to $10^{-2}$~meV. In what follows, we provide a heuristic argument showing why this amplification does not occur and subsequently confirm this conclusion through numerical evaluation.

We consider the projected operator $\mathcal{K}$ on to two localized states at the edges, $|0,i \rangle$ and $|0,i^\prime \rangle$,

\begin{equation}
\label{eq:K1}
\begin{split}
   & \langle 0,i|\mathcal{K}|0,i^\prime \rangle=  \frac{1}{2} \sum_{ \ell,j} {\langle{ 0,i|\frac{l_Bp_x}{\hbar}|\ell,j \rangle}\langle{ \ell,j|\frac{l_Bp_x}{\hbar}|0,i^\prime \rangle} } \times
        \\
        &\times  \left( \frac{\omega_B}{\omega_{LC}+\Delta_{\ell,i,j} } + \frac{\omega_B}{\omega_{LC}+\Delta_{\ell,i^{\prime},j} } \right)~.
\end{split}
\end{equation}

To elucidate the structure of $\mathcal{K}$, we examine the behavior of the matrix elements $\langle 0,i | (l_B p_x/\hbar) | \ell,j \rangle$ and $\langle \ell,j | (l_B p_x/\hbar) | 0,i' \rangle$. As a representative case, we analyze the first matrix element, $\langle 0,i | (l_B p_x/\hbar) | \ell,j \rangle$, which explicitly reads

\begin{equation}
\label{eq:K1}
 \langle{ 0,i|\frac{l_Bp_x}{\hbar}|\ell,j \rangle}=\int d{\bf r}~\psi_{0,i}(\mathbf{r})( -il_B\partial_x)\psi_{\ell,j}(\mathbf{r})~,
\end{equation}
 where $\psi_{\ell,j}(\mathbf{r})=\langle{\bf r}|\ell,j\rangle$ is the wave function of the state $\ket{\ell,j}$.
Two distinct situations arise, depending on whether the intermediate state $\ket{\ell,j}$ is localized or delocalized, corresponding to $\mathcal{K}=\mathcal{K}^{\rm localized}+\mathcal{K}^{\rm extended}$, each leading to qualitatively different contributions.

\begin{enumerate}
\item If the intermediate state $\ket{\ell,j}$ is also exponentially localized, only a limited number of such states overlap with the initial and final ones, all confined within a region of size $l_B$. Being $x_j$ the center of the wavefunction $|\ell,j\rangle$, this integral vanish if $|x_j-x_i|>l_B$. Heuristically, for states separated by less than $l_B$, the matrix element can be approximated as $\langle 0,i | (x/l_B) | \ell,j \rangle \approx 1$, while it exponentially vanishes for more distant states. Consequently, only a fraction $l_B^2/(L_xL_y)$ of all states contributes to the sum, corresponding to an effective number of $N_{\rm L} l_B^2/(L_xL_y) \sim 1$. Thus, $\langle 0,i | \mathcal{K} | 0,i' \rangle$ is of order unity.
If the intermediate state $\ket{\ell,j}$ is also exponentially localized, only a small number of such states overlap with the initial and final ones, all confined within a region of size $l_B$. Let $x_j$ denote the center of the wavefunction $\ket{\ell,j}$; the overlap integral vanishes exponentially when $|x_j - x_i| > l_B$. Conversely, for states separated by less than $l_B$, the matrix element can be approximated as $\langle 0,i | (l_Bp_x/\hbar) | \ell,j \rangle \approx 1$ as the typical variation scale of the wavefuction is given by $l_B$. Hence, only a fraction $l_B^2/(L_xL_y)$ of all states contributes to the sum, corresponding to an effective number $N_{\rm L} l_B^2/(L_xL_y) \sim 1$. Therefore, $\langle 0,i | \mathcal{K}^{\rm localized} | 0,i' \rangle$ is of order unity.

\item We now consider the case where the intermediate states $\ket{\ell,j}$ are extended \cite{halperin_quantized_1982,trugman_localization_1983,ando_localization_1985,arovas_localization_1988}.
To estimate the integral in Eq.~\eqref{eq:K1}, note that the localized wavefunction $\psi_{0,i}(\mathbf{r})$ is confined within an area of order $l_B^2$, giving a normalization factor $1/l_B$.
In contrast, the extended state $\psi_{\ell,j}(\mathbf{r})$ is delocalized over the entire sample area $A=L_xL_y$, contributing a factor $1/\sqrt{L_xL_y}$.
Since $(l_B p_x/\hbar)$ acts over the short scale of the localized wavefunction, we take $(l_B p_x/\hbar)\approx1$.
The spatial integral effectively covers the region where $\psi_{0,i}(\mathbf{r})$ is non-negligible—an area $\sim l_B^2$—thus providing a multiplicative factor $l_B^2$.
Collecting all factors, we obtain $\langle{ 0,i|(l_B p_x/\hbar)|\ell,j \rangle}\approx l_B/\sqrt{L_xL_y}$. 
As an upper bound, we can estimate that the sum over extended states includes all $N_{\rm L}\sim L_xL_y/l_B^2$ available states, leading to $\langle 0,i|\mathcal{K}|0,i'\rangle \sim 1$.
This, however, largely overestimates the actual contribution. According to Ref.~\cite{huo_current_1992}, the number of truly extended states is much smaller, $N_{\rm ext}\ll N_{L}$.
Consequently, although these states mediate long-range hopping, the corresponding contribution scale as
\begin{equation}
\langle 0,i| \mathcal{K}^{\rm extended} |0,i' \rangle \ll 1,
\end{equation}
which is smaller than the contribution from the localized states previously estimated.

\end{enumerate}

\subsection*{Numerical evaluation of vacuum contributions}

\begin{figure}
    \centering
    \begin{overpic}
    [width=\linewidth]{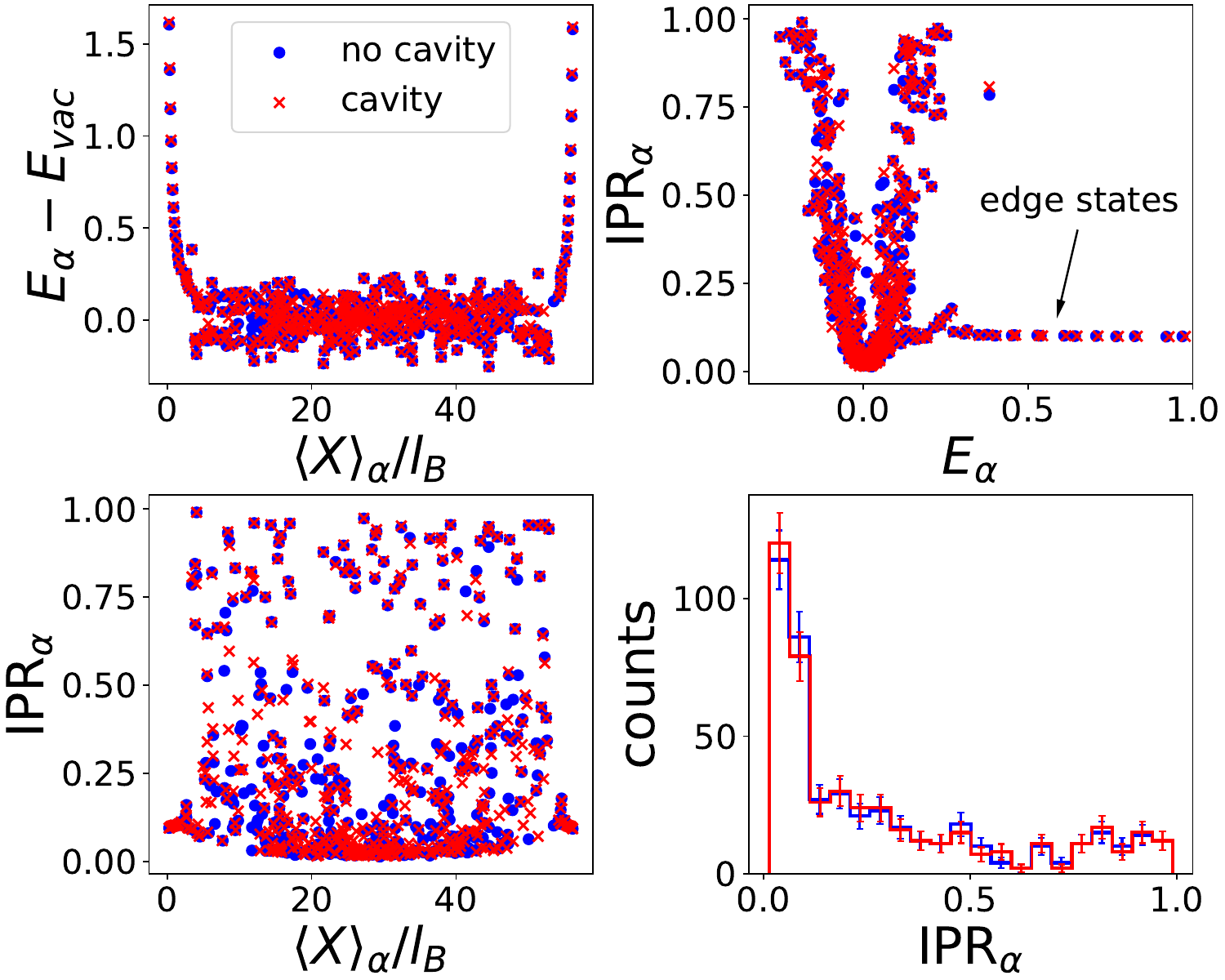}
    \put(20,65){(a)}
    \put(12,20){(b)}
    \put(90,75){(c)}
    \put(90,35){(d)}

    \end{overpic}
    \caption{Effect of cavity-mediated hopping on the single-particle spectrum comparing the spectrum of ${H}_e$  (blu points,  with light-matter coupling $e\mathcal{E}_{vac}l_B=0.$), with the effective Hamiltonian ${H}_{\rm eff}$ of Eq. \eqref{eq:Heff_app} (red points, with light-matter coupling $e\mathcal{E}_{vac}l_B/(\hbar\omega_B)=0.3$). (a) Energy spectrum, removing the diamagnetic-shift of Eq. \eqref{eq:Ha2} $E_{vac}=(e\mathcal{E}_{vac}l_B)e^2/[4\hbar(\omega_{LC}+\omega_B)]$, as a function of average position of the single-particle states $\langle X\rangle_\alpha$. (b) Inverse participation  ratio (IPR) of each eigenstate $\alpha$ as a function of their position. (c) IPR of each eigenstate $\alpha$ as a function of their energy. (d) Statistics of the IPR. In all panels $L_y=50 \,l_B$ for a number of states $M=450$ (hence $L_x\simeq 2\pi l_B^2 M/L_y \simeq 50 \,l_B$); disorder parameters $\xi=l_B$ and $E_{dis}=0.03 \hbar \omega_B$; cavity parameter $\hbar g_e=0.3 \hbar \omega_B $ and $\hbar \omega_{LC}=\hbar\omega_B$. The no-cavity case is solved by keeping $N_{{\rm max}\,\ell}=3$, enough to reach convergence.}
    \label{fig:SI5}
\end{figure}

We now validate the analytical estimation through numerical diagonalization of Eq.~\eqref{eq:Heff_app}.
To construct the effective Hamiltonian $H_{\rm eff}$, we first diagonalize the single-electron Hamiltonian ($H_e$), which includes both disorder and confinement potentials.
The diagonalization is performed by projecting $H_e$ onto the continuous Landau basis, as described in Sec.~\ref{app:Landau_continuous}, and truncating the Hilbert space until convergence of eigenvalues and eigenstates is achieved.
In the weak-disorder limit ($E_{\rm dis} \ll \hbar\omega_B$), the Landau levels $\ell$ remain well defined, though their degeneracy is lifted.
Denoting the resulting eigenstates as $|{\ell,i}\rangle$, we compute numerically all matrix elements entering Eqs.~\eqref{eq:Hpa}–\eqref{eq:K-kernel} and assemble $H_{\rm eff}$ in this single-electron disordered basis ${\ket{\ell,i}}$.

In Fig.~\ref{fig:SI5}, we analyze the eigenstates of Eq.~\eqref{eq:Heff_app}.
Panel (a) shows the energy spectrum—after removing the diamagnetic shift of Eq.~\eqref{eq:Ha2}, as a function of the average position $\langle X \rangle_\alpha$ of each eigenstate. After the removal of this constant diamagnetic shift, the eigenenergies are essentially unaffected by the cavity vacuum contribution.
Panels (b)–(d) characterize the localization properties through the Inverse Participation Ratio (IPR), defined as
\begin{equation}
    {\rm IPR}_\alpha = \int d{\bm r} |\psi_\alpha(\mathbf{r})|^4 ,
\end{equation}

which quantifies the spatial extent of the eigenstate $\psi_\alpha(\mathbf{r})$. Panel (b) displays the IPR versus position, panel (c) versus energy, and panel (d) presents the IPR distribution.
The results show that both the energy spectrum and the degree of localization remain essentially unchanged by the inclusion of the cavity vacuum field. The numerical analysis was performed with parameters chosen larger than the estimated physical values to ensure that any possible effect from the operator $\mathcal{K}$ in Eq.~\eqref{eq:K-kernel} would be clearly visible on the relevant energy scale. As apparent in the Figure, the effective Hamiltonian induced by the cavity vacuum does not seem to have a sizable impact on the disordered eigenstates and their localization properties. 

\begin{figure}
    \centering
    \includegraphics[width=\columnwidth]{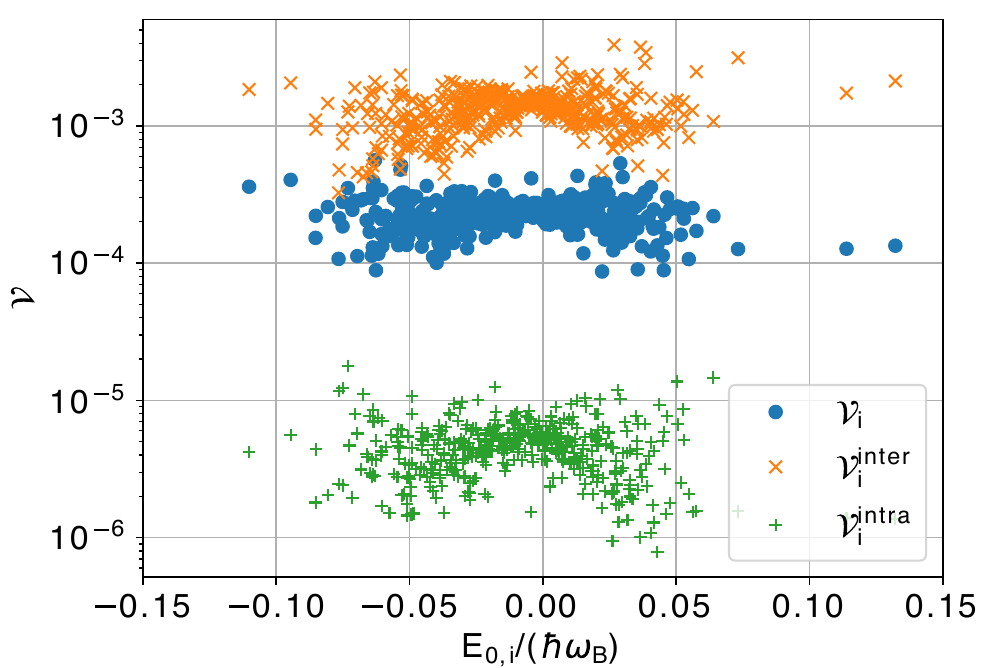}
    \caption{Vacuum spreading estimator $\mathcal{V}_i$ as a function of the respective energy eigenvalue $E_{0,i}$. The simulation has been performed using the Harper-Hofstadter model explained in the SM.
    Parameters: $\alpha=1/25$, $N_x=230$, $N_y=50$, $E_{\rm dis}/(\hbar J) = 0.12$, $\omega_B/J = 4\pi\alpha\approx 0.5$, $\xi_c/l_B\approx 1$. }
    \label{fig:SI6}
\end{figure}

We now analyze Eq.~\eqref{eq:Heff_app} in more detail to understand why the summation over intermediate states in Eq.~\eqref{eq:K-kernel} does not lead to a macroscopic enhancement of vacuum-induced processes in a single particle scenario. Neglecting the constant diamagnetic shift of Eq.~\eqref{eq:Ha2}, we focus on the operator $\mathcal{K}$ in Eq.~\eqref{eq:K-kernel}.

Following Ref.~\cite{ciuti_cavity-mediated_2021}, we introduce the vacuum spreading,
\begin{equation}\label{eq:vacuum_spread}
\mathcal{V}_i = \sum_{i'} \big|\langle 0,i'|\mathcal{K}|0,i\rangle\big|^2 ,
\end{equation}
which quantifies the overall tunneling amplitude between a localized state $\ket{0,i}$ and the other lowest–Landau-level (LLL) states induced by the vacuum field.

To isolate the physical origin of this coupling, we decompose the kinetic operator as
\begin{equation}
\mathcal{K} = \mathcal{K}^{\rm inter} + \mathcal{K}^{\rm intra},
\end{equation}

where $\mathcal{K}^{\rm inter}$ involves virtual transitions to higher Landau levels ($\ell>0$) and $\mathcal{K}^{\rm intra}$ accounts for transitions within the LLL ($\ell=0$) allowed by the presence of disorder.

For the inter–Landau-level contribution, where $\Delta_{\ell,i,j} \approx \omega_B$, we obtain
\begin{equation}\label{eq:K-inter}
\mathcal{K}^{\rm inter} \approx
\frac{\omega_B}{\omega_B+\omega_{LC}}
\frac{1}{l_B^2}
\mathcal{P}x\mathcal{Q}x\mathcal{P},
\end{equation}
where $\mathcal{Q} = \mathbb{I}-\mathcal{P}$ projects onto the excited subspace, and the commutator identity
$p_x = (m/i\hbar)[x,H_e]$
has been used to express $\mathcal{K}^{\rm inter}$ in terms of the position operator.

For the intra–Landau-level component, assuming $\Delta_{\ell,i,j} \ll \omega_B, \omega_{LC}$, we find
\begin{equation}\label{eq:K-intra}
\mathcal{K}^{\rm intra} \approx
\frac{\omega_B}{\omega_{LC}}
\frac{l_B^2}{\hbar^2}
\mathcal{P}p_x\mathcal{P}p_x\mathcal{P}.
\end{equation}

The corresponding contributions to the vacuum spreading are then evaluated as
\begin{align}
\label{eq:inter_contribution}
\mathcal{V}_i^{\rm inter} &=
\sum_{i'} \big|\langle i',0|\mathcal{K}^{\rm inter}|i,0\rangle\big|^2, \\
\label{eq:intra_contribution}
\mathcal{V}_i^{\rm intra} &=
\sum_{i'} \big|\langle i',0|\mathcal{K}^{\rm intra}|i,0\rangle\big|^2.
\end{align}

We perform a numerical simulation by diagonalizing Eq.~\eqref{eq:Heff_app}. The single-electron Hamiltonian $H_e$ is implemented using the Harper–Hofstadter model, with parameters chosen to reproduce the continuum limit. After diagonalizing $H_e$, we construct $H_{\rm para}$ from its exact disordered eigenstates and assemble the full effective Hamiltonian $H_{\rm eff}$, which is then diagonalized.

In Fig.~\ref{fig:SI6}, we show the vacuum spreading [Eq.~\eqref{eq:vacuum_spread}] and the estimators [Eqs.~(\ref{eq:inter_contribution},~\ref{eq:intra_contribution})] as functions of the energy $E_{0,i}$ of each disordered eigenstate in the LLL. For all states these estimators are much below unity.

Figure~\ref{fig:SI6} also compares the individual contributions $\mathcal{V}_i^{\rm inter}$ and $\mathcal{V}_i^{\rm intra}$: the inter-Landau-level component [Eq.~\eqref{eq:inter_contribution}] provides an upper bound to the total vacuum spreading, while the intra-Landau-level term [Eq.~\eqref{eq:intra_contribution}] represents a lower bound.
A code reproducing these numerical results is available in Ref.~\cite{de_bernardis_gihub_2025}.

\section{Screening from layers parallel to the 2DEG}

A common feature of 2DEG in semiconductor heterostructures is the presence of distant parallel doping layers with vertical distances usually around $d_z\simeq100\,$nm (e.g. see SM of Ref. \cite{appugliese_breakdown_2022}). These might provide screening of the image-charge potentials generated by the metallic resonator. While a detailed treatment of the image-charge potential in an experimentally detailed situation is beyond the scope of this work, we want to provide a worse case scenario for the screening of the in-plane image-potential of the metallic split-ring.  
\begin{figure}
    \centering
    \includegraphics[width=0.9\linewidth]{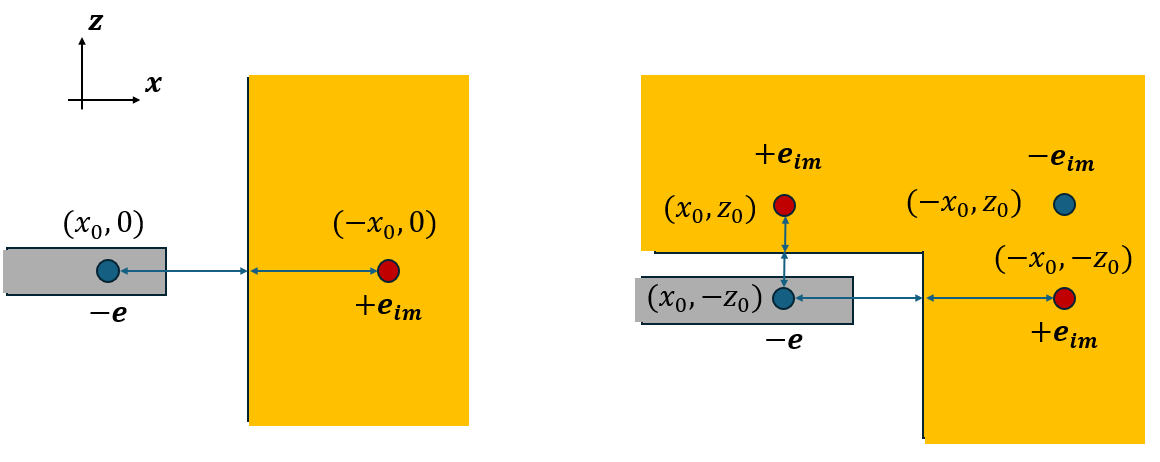}
    \caption{Sketch of the two metallic configurations used to estimate the image-charge potential, the y-direction is not shown as the system is assumed to be translationally invariant there. On the left the simple perpendicular plate also considered in the main text, on the right the corned geometry used to here to put an upper bound on the screening realized by doping layers. Gray area represent the 2DEG layer while yellow areas the metallic regions.}
    \label{fig:SI7}
\end{figure}

We consider a very idealized scenario in which the doping layer is actually replaced by another perfect metallic layer. This puts an upper-bound to the extent of screening provided by the doping layers (which usually consist of immobile charges). The resulting geometry is roughly approximated by a metallic corner as shown in Fig. \ref{fig:SI7}. The electrostatic energy renormalization given by these metallic boundaries is also solvable using the image-charge configuration shown on the right. Assuming the out of plane position is fixed as $z_0=d_z$, we get the following image charge potential renormalization of the charge living in the 2DEG at $(x_0,y_0)$:
\begin{align}
    U_{\mathrm{img}}(x_0)&= -\frac{e^2}{4\pi \epsilon}\left(\frac{1}{2 |x_0|} -\frac{1}{2\sqrt{x_0^2+d_z^2}} +\frac{1}{2d_z}\right) \simeq \nonumber\\
    \simeq&-\frac{e^2}{8\pi \epsilon} \frac{ d_z^2}{2 |x_0|^3}
\end{align}
for $d_z\gg |x_0|$. The important figure of merit is the in-plane gradient of this potential at $x_0=-d_{\mathrm{edge}}$ which gives:
\begin{align}
    |\partial_x U_{\mathrm{img}}(x)| \simeq \frac{e^2}{8\pi \epsilon} \frac{1}{d_{\mathrm{edge}}^2}\frac{3}{2}\frac{d_z^2}{d_{\mathrm{edge}}^2}\;.
\end{align}
For the aforementioned distance $d_z=100\,$nm and $d_{\mathrm{edge}}=200\,$nm; the image charge potential gradient is a factor $\sim 0.38$ less than the case with just a plain metallic layer ($e^2/(8\pi \epsilon d_{\mathrm{edge}}^2)$ used in the main text.

This simple estimate gives a loose sense of the maximum screening strength of in-plane electrostatic potentials by layers parallel to the 2DEG. We stress that the above estimate is highly conservative for a doping layer consisting of immobile donors.

\end{document}